\documentclass[journal,twocolumn]{IEEEtran}
\usepackage[latin1]{inputenc}
\usepackage{amssymb}
\usepackage{amsfonts}
\usepackage{amsthm}
\usepackage{amsmath}
\usepackage{algorithm}
\usepackage{algorithmic}
\usepackage{subeqnarray}
\usepackage{cases}
\usepackage{mathrsfs}
\usepackage{algorithm}
\usepackage{algorithmic}
\usepackage{balance}

\usepackage{graphicx,subfigure,amsmath,amssymb,cite}
\newcommand{\tabincell}[2]{\begin{tabular}{@{}#1@{}}#2\end{tabular}}

\usepackage[dvips]{color}
\usepackage{float}
\ifCLASSINFOpdf
\else
\fi
\begin{document}

\title{\LARGE Multi-RIS Aided 3D Secure Precise Wireless Transmission
}

\author{Tong Shen,~Wenlong Cai,~Yan Lin,~\emph{Member,~IEEE},~Shuo Zhang,~Jinyong Lin,~Feng Shu, \emph{Member,~IEEE} and~Jiangzhou Wang, \emph{Fellow,~IEEE}

\thanks{Tong Shen is with the School of Information and Computer, Anhui Agriculture University, 230036, China. (Email: shentong0107@163.com). }
\thanks{Yan Lin, and Feng Shu are with the School of Electronic and Optical Engineering, Nanjing University of Science and Technology, 210094, CHINA. (Email: yanlin@njust.edu.cn, shufeng0101@163.com). }
\thanks{Wenlong Cai,~Shuo Zhang and Jinyong Lin are with the National Key Laboratory of Science and Technology on Aerospace Intelligence Control, Beijing Aerospace Automatic Control Institute, Beijing, 100854, China (E-mail: caiwenlon@buaa.edu.cn, gcshuo@163.com, ljiny3771@sina.com).}
\thanks{Feng Shu is with School of Information and Communication Engineering, HaiNan University, HaiNan, 570228, China(E-mail: shufeng0101@163.com).}
\thanks{Jiangzhou Wang is with the School of Engineering and Digital Arts, University of Kent, Canterbury CT2 7NT, U.K. (Email: J.Z.Wang@kent.ac.uk.)}}

\maketitle

\begin{abstract}
 In this paper, multiple reconfigurable intelligent surfaces (RIS) aided secure precise wireless transmission (SPWT) schemes are proposed in the three-dimensional (3D) wireless communication scenario. Unavailable direct path channels from transmitter to receivers are considered when the direct paths are obstructed by obstacles. Then, multiple RISs are utilized to achieve SPWT through the reflection path among transmitter, RISs and receivers in order to enhance the communication performance and energy efficiency simultaneously. First, a maximum-signal-to-interference-and-noise ratio (MSINR) scheme is proposed in a single user scenario. Then, the multi-user scenario is considered where the illegitimate users are regarded as eavesdroppers. A maximum-secrecy-rate (MSR) scheme and a maximum-signal-to-leakage-and-noise ratio (MSLNR) are proposed. The former achieves a better secrecy rate (SR) performance but incurs a higher complexity. The latter has a lower complexity than the MSR scheme with an SR performance loss. Simulation results show that both single-user scheme and multi-user scheme can achieve SPWT which transmits confidential message precisely to location of desired users. Moreover, MSLNR scheme has a lower complexity than the MSR scheme, while the SR performance is close to that of the MSR scheme.
\end{abstract}
\section{Introduction}
As a promising physical layer security (PLS) technique in future 6G communications, directional modulation (DM) \cite{Babakhani2008,Babakhani2009,Daly2009Directional,Daly2010Directional} has attracted ever-growing interests from industry and academic due to its unique characteristic, which projects modulated confidential signal into a predetermined spatial direction while makes the constellations of the signal in other directions distorted simultaneously. By contrast, the traditional communication methods based on cryptographic techniques need additional secure channels for private key exchanging and thus may be not applicable in the mobile Internet. Benefiting from the key-less PLS technique \cite{Wang2012Distributed,ChenX2017,Zou2016Relay,JZhao2019,Guo2017Exploiting,zhouTWC2022}, DM can significantly decrease the probability of confidential signal being eavesdropped on and gradually becomes an alternative technology to do encryption in the higher layer.

With a substantial progression of PLS in various aspects, such as massive multi-input-multi-output (MIMO) technique \cite{Wu2017Secure}\cite{Ni2019} and artificial noise (AN) aided secure transmission \cite{Yan2016Artificial}\cite{Zhao2016Anti}, DM has also been rapidy developed by integrating the merits of these PLS techniques. In \cite{Daly2009Directional} and \cite{Daly2010Directional}, the authors proposed a DM signal construction method by shifting the phase of each antenna element which is implemented on the radio frequency (RF) front-end but requires high speed RF switches and high complexity. To overcome those disadvantages, \textit{Hu et al.} in \cite{Hu2016} developed an AN-aided robust baseband DM scheme by considering the estimation errors on direction angles. As a further step, \textit{Shu et al.} in \cite{Shu2016DM} considered multi-user broadcasting scenarios and proposed a leakage-based robust DM scheme. Moreover, in \cite{2019Optimal}, \textit{Lu et al.} put forward a power allocation strategy for transmitting confidential signal and AN, which significantly increases the secrecy performance.

However, DM can only guarantee the security in a specific direction rather than a specific distance, which means that the eavesdroppers can intercept confidential message at the same direction but different distance \cite{book2009}. When the eavesdroppers are located on the same direction as desired receivers, the secure transmission may face with a great challenge. To guarantee a secure transmission in the aforementioned scenario, researchers investigated the combination of PLS and DM \cite{Shu2020DM,Qin2018,Han1} and started the study of secure precise wireless transmission (SPWT) in the two dimensional scenario, which aims to transmit confidential message to a specific direction and distance. Specifically, in \cite{Sammartino2013} and \cite{Wang2015}, the authors proposed a linear frequency diverse array (LFDA) to address the problem that DM can not guarantee the secure transmission in different distance. With the LFDA, the produced beam-pattern is controlled by both direction and distance, thus secure transmission can be guaranteed in the scenario where the desired user and eavesdropper are located in the same direction but different distance. Nevertheless, LFDA scheme has its weakness that the direction and distance of the produced beam-pattern achieved by LFDA are coupled. This means that multiple direction and range pairs may exist at locations where eavesdroppers can receive the identical signal as the desired users, which may still cause security problems. In \cite{liu2017random,Hu2017SPWT,Shu2018SPWT}, the authors proposed a random frequency diverse array (RFDA) scheme, which has a property that can decouple the correlation between the direction and distance. In addition, two random subcarrier selection methods were proposed in \cite{Shen2019}  to achieve SPWT for each modulated orthogonal frequency division multiplexing (OFDM) symbol \cite{Zhu2009Chunk}\cite{Zhu2012Chunk}, which makes SPWT more practical for real scenarios. In \cite{Shen2020}, the authors proposed a hybrid digital and analog (HDB) beamforming scheme which significantly reduces the complexity of SPWT, while reduces the RF-chain circuit complexity in medium-scale and large-scale systems.

Due to the high requirements on the specific direction and distance, SPWT is usually applied in light of sight (LOS) channel scenarios, such as unmanned aerial vehicle (UAV) communication, suburban mobile communication and satellite communication. However, the direct path from the transmitter to the receiver may be obstructed by some obstacles, such as high buildings, mountains and trees. This means that the LOS channel between transmitter and receiver could be unavailable. Thus, the technology of reconfigurable intelligent surface (RIS)\cite{2017IRS}\cite{panCM2021} is introduced to achieve SPWT, since it can reflect the confidential signal and provide an extra transmission path. In other words, RIS is regraded as a relay but more practical since an RIS consists of a number of low-cost passive elements with adjustable phase shifts. Different from amplify-and forward (AF) relay, RIS only reflects the received signal as a passive array where additional power consumption can be avoided. Moreover, a full-duplex modeled RIS system has a higher spectral efficiency than a AF relay modeled transmission system. In \cite{Wu2019}, the authors proposed an enhanced RIS-aided wireless network via joint active and passive beamforming. Then, the authors in \cite{Wu2020Tcom} studied a more practical RIS-aided wireless network, where an RIS is deployed with only a finite number of phase shifts at each element. To transmit both power and information, \textit{Wu et al.} in \cite{Wu2020WCL} proposed a new simultaneous wireless information and power transfer system with the aid of the RIS technology. Moreover, a RIS assisted secure wireless communications with multiple-transmit and multiple-receive antennas was shown in \cite{Jiang2020}. To the best of our knowledge, most recent works related on RIS transmission are based on a single equipped RIS. Since the receiver's position may change frequently to different locations, the direct path between a single RIS and receiver may also be unavailable which may lead to a failure transmission.

As aforementioned, due to the LOS direct path requirement of SPWT, a more reliable secure transmission scheme has to be designed. RIS-aided SPWT is an alternative scheme to overcome the problem that the LOS direct path may be unavailable. In this paper, we combine SPWT with RIS in a direct path unavailable scenario and propose three multi-RISs-aided SPWT schemes. Moreover, a three dimensional (3D) scenario is considered which has one more pitch angle dimension than that of two dimensional (2D) scenario (distance dimension and azimuth angle dimension only). Then, with the goal of maximizing the signal-to-interference-and-noise ratio (SINR) or secrecy rate (SR) at the location of desired users, confidential message energy can be conserved by the desired users while other users can only receive a constellation distorted signal with low useful signal power. Our main contributions are summarized as follows.
\begin{enumerate}
 \item  Considering multi-RIS, a 3D SPWT system with rectangular transmit antenna array is proposed. Due to the fact that the RIS consists of a number of low cost passive elements with adjustable phase shifts, it can significantly reduce the complexity and increase the energy efficiency compared with AF relay. More importantly, the system designs the phase shifts for the part of RISs which has available LOS channel to the user. Compared with single-RIS system which can not guarantee that there exist a LOS direct path between the RIS and the user, multi-RIS system is more practical for real scenarios. Then, a maximum-signal-to-noise ratio (MSINR) scheme is proposed with single user which guarantees the desired user receiving the maximum confidential signal power, while at the same time, the artificial noise is projected on the null-space of the desired user to interfere the location-unknown and potential existed eavesdroppers.

 \item  Considering multi-user scenarios where the part of undesired users are regarded as eavesdroppers, a maximum-secrecy-rate (MSR) scheme is proposed to send confidential message to the desired users and protect the privacy information from eavesdropper at the same time. In this scenario, this scheme is achieved by maximizing the secrecy capacity of the minimal user. The MSR algorithm provides the optimal solutions of beamforming vector, AN vector in transmitter and the phase-shifting in the RISs with an extremely high complexity.

 \item  To reduce the high complexity of MSR scheme, a low-complexity maximum-signal-to-interference-and-noise ratio (MSLNR) scheme is proposed by maximizing the ratio of the confidential signal power received by desired users to that received by eavesdroppers. Numerical results show that this scheme gives a suboptimal solution with a significantly reduced complexity, while the secrecy rate performance of the MSLNR scheme is close to that of the MSR scheme.
\end{enumerate}


The remainder of this paper is organized as follows. In Section II, we propose the RISs-aided 3-D SPWT system model. Then, a MSINR scheme for single user scenario is presented in Section III. In Section IV, a high performance MSR scheme and a low-complexity MSLNR scheme are proposed for multi-user scenario. The performance of the proposed method is evaluated in Section V. Finally, the conclusions are drawn in Section V.

$Notations$: Throughout the paper, matrices, vectors, and scalars are denoted by letters of bold upper case, bold lower case, and lower case, respectively. Signs $(\cdot)^*$, $(\cdot)^T$, $(\cdot)^H$, and $(\cdot)^{-1}$ denote conjugate, matrix transpose, conjugate transpose, and Moore-Penrose inverse, respectively. The symbol $\mathbf{I}_K$ denotes the $K\times K$ identity matrix.
\section{System Model}
\begin{figure}[t]
\centering
\includegraphics[width=0.50\textwidth]{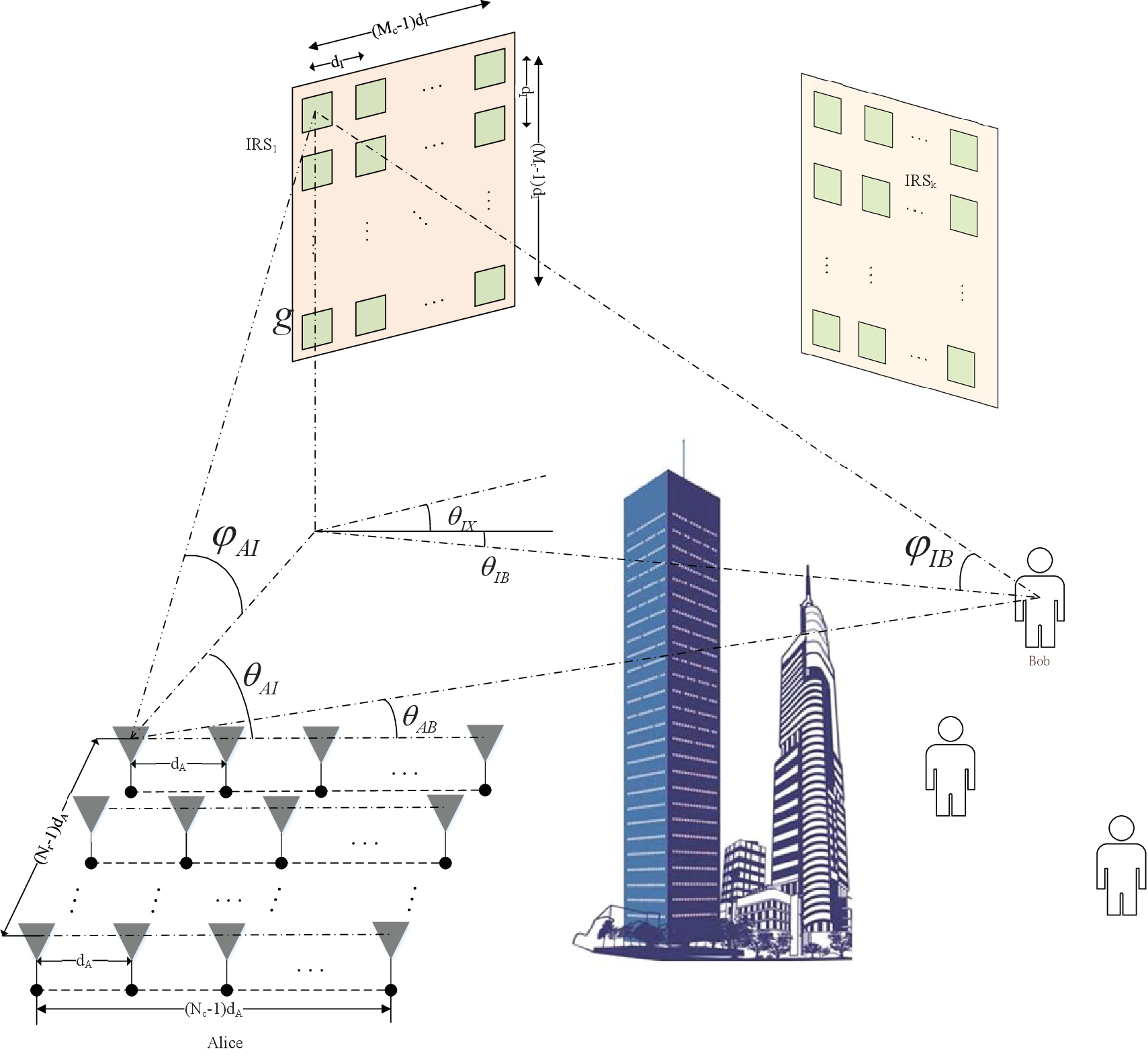}\\
\caption{Schematic diagram of the proposed scheme.}\label{system}
\end{figure}

 As shown in Fig. \ref{system}, consider a multi-RIS-aided SPWT system where the direct path from the transmitter Alice to the receiver Bob is blocked by high-rise buildings. The transmitter Alice is a rectangular antenna array with the antenna number of $N=N_r \times N_c$ and the antenna spacing is denoted as $d_A$. The $K$ RISs with a rectangular array having $M=M_r \times M_c$ elements are equipped on the surface of high-rise building. Assuming that all of the RISs are perpendicular to the ground, the element spacing and height of each RIS are denoted as $d_I$ and $g$, respectively. Without loss of generation, we assume the transmitter Alice and the users are on the ground at the same height. Due to the large path loss, the signal power reflected by the RIS two or more times is negligible. Therefore, it is assumed that the channels from Alice to RIS and RIS to Bob are light of sight (LOS) channel, while the direct path channel from Alice to Bob is unavailable.

 Taking the first antenna of Alice as the reference element, the angle between the antenna array of Alice and the projection of RIS on the ground is denoted as $\theta_{IX}$, the azimuth angle and the pitch angle between Alice and the RIS are denoted as $\theta_{AI}$ and $\varphi_{AI}$, respectively. $\theta_{IB}$ and $\varphi_{IB}$ denote the azimuth angle and the pitch angle between the RIS and Bob, respectively. Moreover, we denote $r_{AI}$, $r_{IB}$ and $r_{AB}$ as the distance from Alice to RIS, the distance from RIS to Bob, and the distance from Alice to Bob, respectively.

 In general, we assume that the location of the RISs and Bob are known by Alice. Assuming that the reference antenna of Alice as the origin, the row and column of rectangular antenna array are regarded as X-axis and Y-axis, respectively. Then we establish a rectangular coordinate system. The locations of Bob and $k$-th RIS are expressed as $(x_B,y_B,0)$ and $(x_{Ik},y_{Ik},g_k)$, respectively. The values of $\theta_{IX}$, $(x_B,y_B,0)$ and $(x_{I_k},y_{I_k},g_k), \forall k=1,2,\ldots,K$ are known in advance.

 According to the location coordinating information of Bob and RIS, the azimuth angles and pitch angles can be obtained by trigonometric function operations which are expressed as
 \begin{align}\label{AIk}
&\theta_{AI_k}=\arctan \frac{y_{I_k}}{x_{I_k}},\\
&\varphi_{AI_k}=\arctan \frac{g_k}{\sqrt{x^2_{I_k}+y^2_{I_k}}},
 \end{align}
 and
 \begin{align}\label{IkB}
&\theta_{I_kB}=\arctan \frac{y_{I_k}-y_b}{x_B-x_{I_k}},\\
&\varphi_{I_kB}=\arctan \frac{g_k}{\sqrt{(x_{I_k-x_B)^2}+(y_{I_k}-y_B)^2}}.
\end{align}
where $\theta_{AI_k}$ and $\varphi_{AI_k}$ denote the azimuth angle and pitch angle between Alice and the $k$-th RIS, respectively. $\theta_{I_kB}$, and $\varphi_{I_kB}$ denote the azimuth angle and pitch angle between Bob and the $k$-th RIS, respectively. The distance from Alice to the $k$-th RIS (the first row first column element) $r_{AI_k}$ and the distance from $k$-th RIS to Bob $r_{I_kB}$ can be expressed as
\begin{align}\label{rAk}
r_{AI_k}=\sqrt{x^2_{I_k}+y^2_{I_k}+g^2_k},
\end{align}
and
\begin{align}\label{rkB}
r_{I_kB}=\sqrt{(x_{I_k}-x_B)^2+(y_{I_k}^2-y_B)+g^2_k},
\end{align}
respectively. Then, the angle and distance information are obtained which can facilitate the next calculations.

In the following, the channel steering vectors between Alice to the $k$-th RIS and the $k$-th RIS to Bob are derived. First, according to SPWT system, random-subcarrier-selection (RSCS) is applied in transmitter's antennas. Then, according to \cite{Shen2019}, the subcarrier frequency allocated to the $n$-th element of transmitter antenna array is given by
\begin{align}\label{fn}
f_n=f_c+\eta_n \Delta f,
\end{align}
where $f_c$ denotes the reference frequency, $\Delta f$ denotes the sub-channel bandwidth and $\eta_n\in \{0,1,\cdots,N_s-1\}$ denotes the chosen subcarrier index. Note that the total bandwidth $B=N_s \Delta f$. It is assumed that the total bandwidth is far less than the reference frequency, i.e. satisfying $N_s \Delta f \ll f_c$.

Then, the phase of the received signal from the reference element of Alice (i.e. the first element with the reference frequency $f_c$) is given by
\begin{align}\label{psiA0}
\psi_{0}=2\pi f_c \frac{r_A}{c},
\end{align}
where $c$ is the light speed, $r_A$ denotes the distance between Alice and receiver (the $k$-th RIS). Likewise, the phase of the received signal from the $n$-th element (i.e., the $n_r$-th row and $n_c$-th column element) of Alice can be expressed as
\begin{align}\label{psiAn}
\psi_{{n_r,n_c}}=&2\pi f_{n_r,n_c}\cdot \nonumber\\
&\frac{r_A+(n_r-1)d_A\cos \beta_A-(n_c-1)d_A\cos \gamma_A}{c},
\end{align}
where $\beta_A$ denotes the angle between the receiver and the columns of the antenna array of Alice, which can be expressed as
\begin{align}\label{betaA}
\beta_A=\arccos(\cos \varphi_A \sin \theta_A).
\end{align}
And $\gamma_A$ is the angle between the receiver and the rows of the antenna array of Alice, which is given by
\begin{align}\label{gammaA}
\gamma_A=\arccos(\cos \varphi_A \cos \theta_A),
\end{align}
where $\varphi_A$ and $\theta_A$ are the receiver's pitch angle and azimuth angle related to Alice, respectively.
Then, the phase shift of the $n_r$-th row and the $n_c$-th column element at Alice relative to the reference element is given by
\begin{align}\label{psiAn}
\Psi_{{n_r,n_c}}=\psi_{k_{n_r,n_c}}-\psi_{k_0}.
\end{align}
Then the normalized steering vector from Alice to a specific location with $(r,\theta,\varphi)$ is given by
\begin{align}\label{hA}
\mathbf{h}_A=\frac{1}{\sqrt{N}}&[e^{j\Psi_{{1,1}}},e^{j\Psi_{{1,2}}},\cdots,e^{j\Psi_{{N_c,1}}},\nonumber\\
&e^{j\Psi_{{2,1}}},\ldots,e^{j\Psi_{{N_c,N_r}}}].
\end{align}

Similarly, the phase shift of the reflected signal at the $m_r$-th row and the $m_c$-th column element of $k$-th RIS relative to the reference element (i.e. the first element of the RIS) is given by
\begin{align}\label{psiI}
\Phi_{{m_r,m_c}}(f_{n_r,n_c})&=\frac{2\pi f_{n_r,n_c}}{c}( r_I+\Delta r_{{m_r,m_c}}(n_r,n_c) \nonumber\\
&-(m_r-1)d_I\cos \beta_{I_k}+(m_c-1)d_I\cos \gamma_{I_k}),
\end{align}
where $f_{n_r,n_c}$ denotes the reflected signal frequency. $\Delta r_{{m_r,m_c}}(n_r,n_c)$ denotes the distance difference related to the $n_r$-th row and the $n_c$-th column element of Alice which can be expressed as
\begin{align}\label{deltar}
\Delta r_{{m_r,m_c}}(n_r,n_c)=&(m_c-1)d_I\cos \varphi_{AI_k} \cos (\theta_{AI_k}-\theta_{I_kX})\nonumber\\
&-(m_r-1)d_I\sin \varphi_{AI_k}.
\end{align}

Note that in (\ref{psiI}), $\beta_{I_k}$  denotes the angle between the column of the $k$-th RIS and the receiver (i.e. desired users), and $\gamma_I$ denotes the angle between the row of RIS and the receiver, which are different from $\beta_A$ and $\gamma_A$ since the RIS has its placement angle $\theta_{IX}$. The expressions of $\beta_I$ and $\gamma_I$ are represented as
\begin{align}\label{betaI}
\beta_{I_k}=\arccos\left(\cos \varphi_{I_k} \sin (\theta_{I_k}+\theta_{I_kX})\right),
\end{align}
and
\begin{align}\label{gammaI}
\gamma_{I_k}=\arccos\left(\cos \varphi_{I_k} \cos (\theta_{I_k}+\theta_{I_kX})\right),
\end{align}
where $\varphi_{I_k}$ is the pitch angle between the $k$-th RIS and the receiver, and $\theta_{I_k}$ is the azimuth angle between the $k$-th RIS and the receiver.

Hence, the steering vector from the $k$-th RIS to the receiver can be expressed as
\begin{align}\label{hIk}
\mathbf{h}_{I_k}(f)=&[1,e^{j\Phi_{{1,2}}(f)},\ldots,e^{j\Phi_{{M_c,1}}(f)},\nonumber\\
&e^{j\Phi_{{2,1}}(f)},\ldots,e^{j\Phi_{{M_r,M_c}}(f)}]^T.
\end{align}

Considering the security problem of SPWT system, AN is used in beamforming to interfere eavesdroppers. Then, the transmitted signal can be expressed as
\begin{align}\label{s}
\mathbf{s}=\sqrt{\alpha P_s}\mathbf{v}x+\sqrt{(1-\alpha)P_s}\mathbf{w}.
\end{align}
where $\alpha$ denotes the power allocation factor which allocates the power of transmitted confidential signal and AN, which satisfies $0\leq\alpha\leq 1$. $P_s$ is the transmit power of Alice. $x$ is a transmitted complex digital modulation symbol with $\mathbb{E}[{|x|^2}]=1$. In addition, $\mathbf{v}\in \mathbb{C}^{N\times 1}$ denotes the beamforming vector with $\|\mathbf{v}\|^2=1$ and $\mathbf{w}\in \mathbb{C}^{N\times 1}$ is the artificial noise with $\|\mathbf{w}\|^2=1$.

Next, by substituting the location information of the $k$-th RIS $(r_{AI_k},\theta_{AI_k},\varphi_{AI_k})$ into (\ref{hA}), and the Bob's location information $(r_{I_kB},\theta_{I_kB},\varphi_{I_kB})$ relative to the $k$-th RIS into (\ref{hIk}), respectively, we can obtain the steering vector $\mathbf{h}_{A_k}$ from Alice to the $k$-th RIS and the steering vector $\mathbf{h}_{I_kB}$ from the $k$-th RIS to Bob , respectively. Then, the received signal reflected from $K$ RISs to Bob can be expressed as
\begin{align}\label{yb}
y_{AIB}&=\sum \limits_{k=1}^K \rho_{AI_kB}\mathbf{h}^H_{A_k} \mathrm{diag}(\mathbf{s})\cdot\nonumber\\
&[\mathbf{h}_{I_kB}(f_{1,1}),\mathbf{h}_{I_kB}(f_{1,2}),\ldots,\mathbf{h}_{I_kB}(f_{N_r,N_c})]^H\mathbf{\Theta}_k+n_B,
\end{align}
where $\rho_{AI_kB}$ denotes the path loss through the reflection path and is proportional to $\frac{1}{(r_{AI_k}\times r_{I_kB})^2}$. $n_B$ is the additive white Gaussian noise (AWGN) at Bob with the distribution $\mathcal{CN}\sim(0,\sigma^2)$. $\Theta$ represents the phase-shift matrix of RIS which is expressed as
\begin{align}\label{Theta}
\mathbf{\Theta}_k=\left[ e^{j\varphi_{k_{1,1}}},e^{j\varphi_{k_{1,2}}},\ldots ,e^{j\varphi_{k_{m_r,m_c}}} \right]^T.
\end{align}
Defining $\mathbf{H}_{I_kB}(\mathbf{f})=[\mathbf{h}_{I_kB}(f_{1,1}),\ldots,\mathbf{h}_{I_kB}(f_{N_r,N_c})]$, (\ref{yb}) can be rewritten as
\begin{align}\label{yb1}
y_{AIB}&=\sum \limits_{k=1}^K [\rho_{AI_kB}\sqrt{\alpha P_s}\mathbf{h}^H_{A_k} \mathrm{diag}(\mathbf{v})\mathbf{H}^H_{I_kB}(\mathbf{f})\mathbf{\Theta}_k x\nonumber\\
&+\rho_{AI_kB}\sqrt{(1-\alpha) P_s}\mathbf{h}^H_{A_k} \mathrm{diag}(\mathbf{w})\mathbf{H}^H_{I_kB}(\mathbf{f})\mathbf{\Theta}_k]
+n_B.
\end{align}
Accordingly, the SINR of Bob is given by (\ref{SINRb}).
\begin{figure*}
\begin{align}\label{SINRb}
\mathrm{SINR}_B=\frac{|\sum \limits_{k=1}^K \rho_{AI_kB}\sqrt{\alpha P_s}\mathbf{h}^H_{A_k} \mathrm{diag}(\mathbf{v})\mathbf{H}^H_{I_kB}(\mathbf{f})\mathbf{\Theta}_k|^2}{|\sum \limits_{k=1}^K\rho_{AI_kB}\sqrt{(1-\alpha) P_s}\mathbf{h}^H_{A_k} \mathrm{diag}(\mathbf{w})\mathbf{H}^H_{I_kB}(\mathbf{f})\mathbf{\Theta}_k|^2+\sigma^2}
\end{align}
\hrulefill
\end{figure*}

\section{Proposed MSINR scheme in single-user system}
In this section, we consider a single-user scenario to achieve SPWT by optimizing both the beamforming in Alice and the phase-shifting in RISs. Since the eavesdroppers' locations are usually unknown by Alice, their channel state information (CSI) is unavailable, we do not consider the eavesdroppers in this scenario and the eavesdroppers scenario is discussed in next section. Thus, we resort to maximize the SINR at Bob with a given power allocation factor $\alpha$. As a result, AN can be projected in the null space of Bob, and only potential eavesdroppers will be influenced by AN.

In this case, the optimization problem is transformed to
\begin{subequations}\label{P1}
\begin{align}
&\mathop {{\max}}\limits_{\mathbf{v},\mathbf{w},\Theta}  ~\mathrm{SINR}_B \label{P1_1}\\
&~\mathrm{s}.\mathrm{t}.~~\|\mathbf{v}\|^2=1,\|\mathbf{w}\|^2=1, \label{P1_3} \\
&~~~~~~~(\ref{Theta}),\forall k=1,2,\ldots,K.\label{P1_4}
\end{align}
\end{subequations}

 Let us define auxiliary variables $\Omega_k=diag(\mathbf{h}^H_{A_k})$ and $\Upsilon_k=\rho_{AI_kB}\sqrt{ P_s}$. Then, according to the matrix transformation, we can rewrite $\mathbf{h}^H_{A_k} \mathrm{diag}(\mathbf{v})$ and $\mathbf{h}^H_{A_k} \mathrm{diag}(\mathbf{w})$ to $\mathbf{v}^T\Omega_k$ and $\mathbf{w}^T\Omega_k$, respectively.  As a result, problem (\ref{P1}) can be rewritten as
\begin{subequations}\label{P2}
\begin{align}
&\mathop {{\max}}\limits_{\mathbf{v},\mathbf{w},\Theta}  ~\frac{|\sqrt{\alpha}\mathbf{v}^T\sum \limits_{k=1}^K \Upsilon_k \Omega_k \mathbf{H}^H_{I_kB}(\mathbf{f})\mathbf{\Theta}_k|^2}{|\sqrt{1-\alpha}\mathbf{w}^T\sum \limits_{k=1}^K\Upsilon_k\Omega_k \mathbf{H}^H_{I_kB}(\mathbf{f})\mathbf{\Theta}_k|^2+\sigma^2}\label{P2_1}\\
&~\mathrm{s}.\mathrm{t}.~~\|\mathbf{v}\|^2=1,\|\mathbf{w}\|^2=1  \label{P2_3}\\
&~~~~~~~(\ref{Theta}),\forall k=1,2,\ldots,K.\label{P2_4}
\end{align}
\end{subequations}
In (\ref{P2_1}), another auxiliary variable is defined as $\Gamma =\sum \limits_{k=1}^K \Upsilon_k \Omega_k \mathbf{H}^H_{I_kB}(\mathbf{f})\mathbf{\Theta}_k$, $\Gamma \in \mathbb{C}(N\times 1)$ to simplify the objective function. For any given $\Theta_k (\forall k=1,2,\ldots,K)$, it is known that the optimal transmit beamforming solution to (\ref{P2}) is
\begin{align}\label{vopt}
\mathbf{v}_{opt}=\frac{\Gamma^*}{\|\Gamma\|},
\end{align}
and the optimal AN vector is
\begin{align}\label{wopt}
\mathbf{w}_{opt}&=[\frac{\mathbf{z}[\|\Gamma\|^2\mathrm{I}_N-\Gamma \Gamma^H]}{\|\mathbf{z}[\|\Gamma\|^2\mathrm{I}_N-\Gamma \Gamma^H]\|}]^T\nonumber\\
&=\frac{[\|\Gamma\|^2\mathrm{I}_N-\Gamma^* \Gamma^T]\mathbf{z}^T}{\|\mathbf{z}[\|\Gamma\|^2\mathrm{I}_N-\Gamma \Gamma^H]\|},
\end{align}
where $\mathbf{z}$ consists of $N$ i.i.d circularly-symmetric complex Gaussian random variables with zero-mean and unit-variance, i.e., $\mathbf{\mathbf{z}}\sim \mathcal{CN}(0,\mathbf{I}_N)$.

Substituting (\ref{vopt}) and (\ref{wopt}) into (\ref{SINRb}), the optimal SINR at Bob is given by
\begin{align}\label{SINRopt1}
\mathrm{SINR}_{B}=\frac{|\sqrt{\alpha}\sum \limits_{k=1}^K  \mathbf{v}^T\Upsilon_k \Omega_k \mathbf{H}^H_{I_kB}(\mathbf{f})\mathbf{\Theta}_k|^2}{\sigma^2}.
\end{align}

Accordingly, the optimization problem (\ref{P2}) is equivalent to
\begin{subequations}\label{P3}
\begin{align}
&\mathop {{\max}}\limits_{\Theta}  ~|\sum \limits_{k=1}^K\mathbf{v}^T \Upsilon_k \Omega_k \mathbf{H}^H_{I_kB}(\mathbf{f})\mathbf{\Theta}_k|^2\label{P3_1}\\
&~\mathrm{s}.\mathrm{t}.~~(\ref{Theta}),\forall k=1,2,\ldots,K.\label{P3_2}
\end{align}
\end{subequations}

Since $\Upsilon_k$, $\Omega_k$ and $\mathbf{H}^H_{I_kB}(\mathbf{f})$ in (\ref{P3_1}) are known, and $\mathbf{v}^T\Upsilon_k \Omega_k \mathbf{H}^H_{I_kB}(\mathbf{f})$ is a $1\times M$ vector. Thus, (\ref{P3_1}) can be rewritten as
\begin{align}\label{thetaopt}
&|\sum \limits_{k=1}^K \mathbf{v}^T\Upsilon_k \Omega_k \mathbf{H}^H_{I_kB}(\mathbf{f})\mathbf{\Theta}_k|^2\nonumber\\
&=|\sum \limits_{k=1}^K (\sum \limits_{m=1}^M A_{k_{m}}e^{j\gamma_{k_{m}}}e^{j\varphi_{k_m}})|^2,
\end{align}
where $A_{k_{m}}$ is a  positive real number which denotes the amplitude of the $m$-th element in $\mathbf{v}^T\Upsilon_k \Omega_k \mathbf{H}^H_{I_kB}(\mathbf{f})$, and $0\leq\gamma_{k_{m,n}}\leq2\pi$ denotes the phase angle of the $m$-th element in $\mathbf{v}^T\Upsilon_k \Omega_k \mathbf{H}^H_{I_kB}(\mathbf{f})$. In addition, the subscript $m$ in (\ref{thetaopt}) is the simplified expression of $(m_r,m_c)$, which satisfies $m=(m_r-1)M_c+m_c,(0\leq m_r\leq M_r,0\leq m_c\leq M_c)$.

According to (\ref{thetaopt}), it is clear to see that when $\sum \limits_{m=1}^M A_{k_{m,n}}e^{j\gamma_{k_{m,n}}}e^{j\varphi_{k_m}}$ for each $k$ has the same phase angle, $|\sum \limits_{k=1}^K (\sum \limits_{m=1}^M A_{k_{m}}e^{j\gamma_{k_{m}}}e^{j\varphi_{k_m}})|^2$ achieves the maximum value. Thus, let us denote a constant phase angle for each $k$ of  $\sum \limits_{m=1}^M A_{k_{m,n}}e^{j\gamma_{k_{m,n}}}e^{j\varphi_{k_m}}$ to $\varphi_0$, i.e.,
\begin{align}\label{varphi0}
\arg(\sum \limits_{m=1}^M A_{k_{m}}e^{j\gamma_{k_{m}}}e^{j\varphi_{k_m}})=\varphi_0,\forall k=1,2,\ldots,K,
\end{align}
In this case, the optimization problem (\ref{P3}) can be transformed to $K$ sub-optimization problem as
\begin{align}\label{P4}
&\mathop {{\max}}\limits_{\varphi_{k_m}}  ~|\sum \limits_{m=1}^M A_{k_{m}}e^{j\gamma_{k_{m}}}e^{j\varphi_{k_m}}|,\forall k=1,2,\ldots,K.
\end{align}
Next, we have the following inequality:
\begin{align}\label{upperbound}
|\sum \limits_{m=1}^M A_{k_{m}}e^{j\gamma_{k_{m}}}e^{j\varphi_{k_m}}|\leq \sum \limits_{m=1}^M |A_{k_{m}}e^{j\gamma_{k_{m}}}e^{j\varphi_{k_m}}|,
\end{align}
when $\gamma_{k_m}+\varphi_{k_m}=\varphi_0(\forall m=1,2,\ldots,M$), and the inequality (\ref{upperbound}) satisfies the equality condition. Accordingly, the optimal value of $\varphi_{k_m}$ is given by
\begin{align}\label{varphikm}
\varphi_{k_m}=\varphi_0-\gamma_{k_m}.
\end{align}
Then, after substituting $\varphi_{k_m}$ into (\ref{SINRopt1}), the optimal $\mathbf{v}$ and $\Theta_k$ can be obtained within a finite number of iterations. The detailed process is shown in Algorithm 1. To be specific, first, giving an initial value of $\mathbf{\Theta_k}[0],\forall k=1,\ldots,K$, according to (\ref{vopt}), we can obtain $\mathbf{v}[0]$. Next, substituting $\mathbf{v}[0]$ into (\ref{SINRopt1}), $\mathbf{\Theta_k}[1],\forall k=1,\ldots,K$ can be obtained by (\ref{varphikm}). At last, we repeat the above process until convergence, and obtain the optimal solution  $\mathbf{v}_{opt}$ and $\mathbf{\Theta}_{k_{opt}},\forall k=1,\ldots,K$ . Since $\mathrm{SINR}_B(\mathbf{v}[n+1],\mathbf{\Theta}[n+1])\geq \mathrm{SINR}_B(\mathbf{v}[n+1],\mathbf{\Theta}[n])\geq \mathrm{SINR}_B(\mathbf{v}[n],\mathbf{\Theta}[n])$, where$\mathbf{\Theta}[n]={\mathbf{\Theta}_1[n],\ldots,\mathbf{\Theta}_K[n]}$, Algorithm 1 converges.

\begin{algorithm}[t]
\caption{Iteration algorithm for the MSINR scheme}
\label{alg:A}
\begin{algorithmic}[1]
\STATE {Initialize $\mathbf{\Theta_k}[0](\forall k=1,\ldots,K)$ randomly that is feasible to (\ref{P2});}
\STATE {Set $n=0$;}
\REPEAT
\STATE {Obtain $\mathbf{v}[n]$ by (\ref{vopt});}
\STATE {Substitute $\mathbf{v}[n]$ into (\ref{SINRopt1}), and obtain $\mathbf{\Theta_k}[n+1]$ by (\ref{varphikm});}
\STATE {Update n=n+1;}
\UNTIL{convergence}
\STATE {Output the final optimal solution $\mathbf{v}_{opt}=\mathbf{v}[n]$, $\mathbf{\Theta}_{k_{opt}}=\Theta_k[n], \forall k=1,\ldots,K$.}
\end{algorithmic}
\end{algorithm}

Furthermore, it should be noted that, since $\mathbf{w}$ is designed on the null-space of Bob, we have
\begin{align}\label{null-space}
\mathbf{w}^T\sum \limits_{k=1}^K\Upsilon_k\Omega_k \mathbf{H}^H_{I_kB}(\mathbf{f})\mathbf{\Theta}_k=0.
\end{align}
Thus, by substituting $\mathbf{\Theta}_{k_{opt}},\forall k=1,\ldots,K$ into (\ref{wopt}), the optimal solution of $\mathbf{w}$ can be obtained.

\section{Proposed Schemes in multi-user system}
In this section, we consider a multi-user multi-RIS-aided SPWT system. Let $P$ denote the number of desired users which confidential message is transmitted to, and let $Q$ denote the number of undesired users which are not expected to receive the confidential signals. We regard $P$ desired users as legitimate users (Bobs) and $Q$ undesired users as eavesdroppers (Eves), respectively. To protect the confidential signal from eavesdropping by Eves, the optimization problem can be established by maximizing the SR.
\subsection{Proposed MSR scheme}
According to the equation of SINR (\ref{SINRb}) in Section III, the SINR at the $p$-th desired user $\mathrm{SINR}_{b_p}$ can be given by substituting $\rho_{AI_kB_p}$ and $\mathbf{H}^H_{I_kB_p}$ into (\ref{SINRb}). Similarly, the SINR at the $q$-th eavesdropper $\mathrm{SINR}_{E_q}$ can be given by substituting $\rho_{AI_kE_q}$ and $\mathbf{H}^H_{I_kE_q}$ into (\ref{SINRb}). Then, the SR can be expressed as

\begin{align}\label{SR}
\mathrm{SR}=\max \{0,\min \limits_{p,q} \{\log(1+\mathrm{SINR}_{B_p})-\log(1+\mathrm{SINR}_{E_q})\}\}.
\end{align}

Accordingly, the optimization problem of maximizing the SR is constructed as
\begin{subequations}\label{P4}
\begin{align}
&\mathop {{\max}}\limits_{\mathbf{v},\mathbf{w},\Theta,\alpha}  ~\mathrm{SR}\label{P4_1}\\
&~\mathrm{s}.\mathrm{t}.~~\|\mathbf{v}\|^2=1,\|\mathbf{w}\|^2=1  \label{P4_3}\\
&~~~~~~~(\ref{Theta}),\forall k=1,2,\ldots,K.\label{P4_4}
\end{align}
\end{subequations}
However, the objective function in (\ref{P4_1}) is composed of the logarithmic function of the product of fractional quadratic functions, which is non-convex and difficult to tackle. Thus, semi-definite relaxation and first order Taylor expansion are employed in transforming the original problem into a convex problem. The detailed process is illustrated as follows.

First, we can observe that in (\ref{SINRb}, the beamforming vector $\mathbf{v}$, the AN vector $\mathbf{w}$ and the phase shifting in RIS $\Theta$ are coupled, which makes the joint optimization problem very difficult to solve. Therefore, we give an initial value of $\Theta_k,\forall k=1,2,\ldots,K$. Then, let $\Gamma_p =\sum \limits_{k=1}^K \Upsilon_k \Omega_k H^H_{I_kB_p}(\mathbf{f})\mathbf{\Theta}_k$ and $\Gamma_q =\sum \limits_{k=1}^K \Upsilon_k \Omega_k H^H_{I_kE_q}(\mathbf{f})\mathbf{\Theta}_k$. Accordingly, problem (\ref{P4}) can be simplified as
\begin{subequations}\label{P5}
\begin{align}
&\mathop {{\max}}\limits_{\mathbf{v{'}},\mathbf{w{'}}}  ~\min\limits_{p,q} \{\log(1+\frac{|\mathbf{v{'}}^T\Gamma_p|^2}{|\mathbf{w{'}}^T\Gamma_p|^2+\sigma^2})\nonumber\\
&~~~~~~~~~~~~~-\log(1+\frac{|\mathbf{v{'}}^T\Gamma_q|^2}{|\mathbf{w{'}}^T\Gamma_q|^2+\sigma^2})\},\label{P5_1}\\
&~\mathrm{s}.\mathrm{t}.~~\|\mathbf{v^{'}}\|^2+\|\mathbf{w{'}}\|^2=1.\label{P5_2}
\end{align}
\end{subequations}

Note that in (\ref{P5}), power allocation factor $\alpha$ is integrated into $\mathbf{v{'}}$ and $\mathbf{w{'}}$, i.e., $\mathbf{v}=\alpha\mathbf{v{'}}$ and $\mathbf{w{'}}=(1-\alpha)\mathbf{w}$. Thus the power constraint is given as (\ref{P5_2}), and then the optimization variables are reduced to two.

Next, we introduce a set of the exponential variables to substitute the numerators and denominators of the fractions in the objective function in (\ref{P5}), i.e.,
\begin{align}\label{evariable}
&e^{\mu_p}=|\mathbf{v{'}}^T\Gamma_p|^2+|\mathbf{w{'}}^T\Gamma_p|^2+\sigma^2,\\
&e^{\nu_p}=|\mathbf{w{'}}^T\Gamma_p|^2+\sigma^2,\\
&e^{\lambda_q}=|\mathbf{v{'}}^T\Gamma_q|^2+|\mathbf{w{'}}^T\Gamma_q|^2+\sigma^2,\\
&e^{\omega_q}=|\mathbf{w{'}}^T\Gamma_q|^2+\sigma^2.
\end{align}
According to the properties of exponential and logarithmic functions, the problem (\ref{P5}) can be rewritten as
\begin{subequations}\label{P6}
\begin{align}
&\mathop {{\max}}\limits_{\mathbf{v{'}},\mathbf{w{'}}}  ~\min\limits_{p,q} \{\mu_p-\nu_p-\lambda_q+\omega_q\},\label{P6_1}\\
&~\mathrm{s}.\mathrm{t}.~~\|\mathbf{v{'}}\|^2+\mathbf{w{'}}\|^2=1,\label{P6_2}\\
&~~~~~~~|\mathbf{v{'}}^T\Gamma_p|^2+|\mathbf{w{'}}^T\Gamma_p|^2+\sigma^2\geq e^{\mu_p},\label{P6_3}\\
&~~~~~~~|\mathbf{w{'}}^T\Gamma_p|^2+\sigma^2\leq e^{\nu_p},\label{P6_4}\\
&~~~~~~~|\mathbf{v{'}}^T\Gamma_q|^2+|\mathbf{w{'}}^T\Gamma_q|^2+\sigma^2\leq e^{\lambda_q},\label{P6_5}\\
&~~~~~~~|\mathbf{w{'}}^T\Gamma_q|^2+\sigma^2\geq e^{\omega_q}\label{P6_6}.
\end{align}
\end{subequations}

Then, the objective function (\ref{P6_1}) is a concave function because it is a minimum of the affine functions. However, the constraint (\ref{P6_3})-(\ref{P6_6}) are all non-convex. In order to transform these constraints into convex ones, we substitute the positive semi-definite matrix variables $R_v=\mathbf{v{'}}^*\mathbf{v{'}}^T$, $R_w=\mathbf{w{'}}^*\mathbf{w{'}}^T$ and $R_{\Gamma_p}=\Gamma_p\Gamma_p^H$ into (\ref{P6}). Note that $R_v$ and $R_w$ need to satisfy $R_v,R_W\succeq0$, and $\mathrm{rank}(R_v,R_w)=1$. Since the rank-one constraint is still non-convex, we apply the semi-definite relaxation (SDR) to relax this constraint.
As a result, the optimization problem (\ref{P6}) is reduced to
\begin{subequations}\label{P7}
\begin{align}
&\mathop {{\max}}\limits_{R_v,R_w}  ~\min\limits_{p,q} \{\mu_p-\nu_p-\lambda_q+\omega_q\},\label{P7_1}\\
&~\mathrm{s}.\mathrm{t}.~~\mathrm{Tr}(R_v)+\mathrm{Tr}(R_w)=1,\label{P7_2}\\
&~~~~~~~\mathrm{Tr}(R_{\Gamma_p}R_v)+\mathrm{Tr}(R_{\Gamma_p}R_w)+\sigma^2\geq e^{\mu_p},\forall p,\label{P7_3}\\
&~~~~~~~\mathrm{Tr}(R_{\Gamma_p}R_w)+\sigma^2\leq e^{\nu_p},\forall p,\label{P7_4}\\
&~~~~~~~\mathrm{Tr}(R_{\Gamma_q}R_v)+\mathrm{Tr}(R_{\Gamma_q}R_w)+\sigma^2\leq e^{\lambda_q},\forall q,\label{P7_5}\\
&~~~~~~~\mathrm{Tr}(R_{\Gamma_q}R_w)+\sigma^2\geq e^{\omega_q},\forall q,\label{P7_6}\\
&~~~~~~~R_v,R_w\succeq 0 ,\label{P7_7}
\end{align}
\end{subequations}
In this case, the constraints in (\ref{P7_3}) and (\ref{P7_6}) are convex. However, the constraints in (\ref{P7_4}) and (\ref{P7_5}) are still non-convex, thus we linearize $e^{\nu_p}$ and $e^{\lambda_q}$ based on the first-order Taylor approximation as
\begin{align}\label{taylor}
e^{\nu_p}=e^{\bar{\nu_p}}(\nu_p-\bar{\nu_p}+1),\\
e^{\lambda_q}=e^{\bar{\lambda_q}}(\lambda_q-\bar{\lambda_q}+1),
\end{align}
where $\bar{\nu_p}$ and $\bar{\lambda_q}$ are the approximation point, and the approximations are made around
\begin{align}
&\bar{\nu_p}=\ln\left(\mathrm{Tr}(R_{\Gamma_p}R_w)+\sigma^2 \right),\label{appr1_1}\\
&\bar{\lambda_q}=\ln\left(\mathrm{Tr}(R_{\Gamma_q}R_v)+\mathrm{Tr}(R_{\Gamma_q}R_w)+\sigma^2 \right)\label{appr1_2}.
\end{align}
Hence, the problem (\ref{P7}) is eventually transformed as
\begin{subequations}\label{P8}
\begin{align}
&\mathop {{\max}}\limits_{R_v,R_w}  ~\min\limits_{p,q} \{\mu_p-\nu_p-\lambda_q+\omega_q\},\label{P8_1}\\
&~\mathrm{s}.\mathrm{t}.~~\mathrm{Tr}(R_v)+\mathrm{Tr}(R_w)=1,\label{P8_2}\\
&~~~~~~~\mathrm{Tr}(R_{\Gamma_p}R_v)+\mathrm{Tr}(R_{\Gamma_p}R_w)+\sigma^2\geq e^{\mu_p},\forall p,\label{P8_3}\\
&~~~~~~~\mathrm{Tr}(R_{\Gamma_p}R_w)+\sigma^2\leq e^{\bar{\nu_p}}(\nu_p-\bar{\nu_p}+1),\forall p,\label{P8_4}\\
&~~~~~~~\mathrm{Tr}(R_{\Gamma_q}R_v)+\mathrm{Tr}(R_{\Gamma_q}R_w)+\sigma^2\leq \nonumber\\
&~~~~~~~~~~~~~~~~~~~~~~~~~~~~~~~~~~~e^{\bar{\lambda_q}}(\lambda_q-\bar{\lambda_q}+1),\forall q,\label{P8_5}\\
&~~~~~~~\mathrm{Tr}(R_{\Gamma_q}R_w)+\sigma^2\geq e^{\omega_q},\forall q,\label{P8_6}\\
&~~~~~~~R_v,R_w\succeq 0.\label{P8_7}
\end{align}
\end{subequations}

Now, the objective function and all the constraints in (\ref{P8}) are convex, and problem (\ref{P8}) is a convex semi-definite program (SDP), thus it can be optimally solved by existing convex optimization solutions such as CVX \cite{CVX2004}. However, in general the solution of $R_v$ and $R_w$ may not satisfy the relaxed constraint $\mathrm{rank}(R_v,R_w)=1$, which implies that the optimal solution of problem (\ref{P8}) only serves an upper bound of problem (\ref{P6}). Thus, it is necessary to construct a rank-one solution from the obtained higher-rank solution to problem (\ref{P8}). The detailed steps are shown as follows.

1) First, we decompose $R_v$ and $R_w$ as $R_v=U_v\Sigma_vU^H_v$ and $R_v=U_w\Sigma_wU^H_w$, respectively. Herein, $U_v$ and $U_w$ are unitary matrices. $\Sigma_v$ and $\Sigma_w$ are eigenvalue diagonal matrices of $R_v$ and $R_w$, respectively. They have the same size of $N\times N$.

2) Next, we obtain the suboptimal solution of (\ref{P6}) as $\bar{\mathbf{v{'}}}=U_v\Sigma^{\frac{1}{2}}_v\xi_v$ and $\bar{\mathbf{w{'}}}=U_w\Sigma^{\frac{1}{2}}_w\xi_w$, where $\xi_v,\xi_w\sim \mathcal{CN}(0,\mathbf{I}_{N\times 1})$ are $N\times 1$ random vectors.

3) Finally, with sufficient independent generated random vectors $\xi_{v}$ and $\xi_w$, the optimal solution of (\ref{P6}) is approximated as $\bar{\mathbf{v{'}}}$ and $\bar{\mathbf{w{'}}}$ which achieves the highest objective value among all $\xi_v$ and $\xi_w$.

After the solutions of $\mathbf{v{'}}$ and $\mathbf{w{'}}$ are obtained with given $\Theta_k$, the next step is to optimize $\Theta_k$ with the obtained $\mathbf{v{'}}$ and $\mathbf{w{'}}$. The optimization problem can be constructed as
\begin{subequations}\label{P9}
\begin{align}
&\mathop {{\max}}\limits_{\Theta_k}  ~\min\limits_{p,q} \{\log(1+\mathrm{SINR}_{B_p})-\log(1+\mathrm{SINR}_{E_q})\},\label{P9_1}\\
&~\mathrm{s}.\mathrm{t}.~~|\Theta_{k_m}|^2=1,\forall k=1,\ldots,K,\forall m=1,\ldots,M.\label{P9_2}
\end{align}
\end{subequations}
Observing (\ref{SINRb}), all the parameters in $\rho_{AI_kB}\sqrt{\alpha P_s}\mathbf{h}^H_{A_k}\mathrm{diag}(\mathbf{v})\mathbf{H}^H_{I_kB}(\mathbf{f})$ are known, thus for simplicity we denote
\begin{align}\label{hvbp}
\zeta_{kB_p}=\rho_{AI_kB}\sqrt{\alpha P_s}\mathbf{h}^H_{A_k}\mathrm{diag}(\mathbf{v})\mathbf{H}^H_{I_kB}(\mathbf{f}),
\end{align}
similarly, we denote
\begin{align}\label{hwbp}
\eta_{kB_p}=\rho_{AI_kB}\sqrt{\alpha P_s}\mathbf{h}^H_{A_k}\mathrm{diag}(\mathbf{w})\mathbf{H}^H_{I_kB}(\mathbf{f}).
\end{align}
Then, the SINR at the $p$-th desired user is reduced as
\begin{align}\label{SINRbp}
\mathrm{SINR}_{B_p}=\frac{|\sum \limits_{k=1}^K \zeta_{kB_p} \mathbf{\Theta}_k|^2}{|\sum \limits_{k=1}^K\eta_{kB_p}\mathbf{\Theta}_k|^2+\sigma^2}.
\end{align}
Since $\zeta_{kB_p}$ and $\eta_{kB_p}$ are $1\times M$ vectors and $\Theta_k$ is a $M\times 1$ vector, the summation can be rewritten as
\begin{align}\label{sums1}
\sum \limits_{k=1}^K \zeta_{kB_p} \mathbf{\Theta}_k=\mathbf{O}_{B_p}\mathbf{\Theta}
\end{align}
and
\begin{align}\label{sums2}
\sum \limits_{k=1}^K \eta_{kB_p} \mathbf{\Theta}_k=\mathbf{U}_{B_p}\mathbf{\Theta}
\end{align}
where $\mathbf{O}_{B_p}$, $\mathbf{U}_{B_p}$ and $\Theta$ are the composite vector of $\zeta_{kB_p}$, $\eta_{kB_p}$ and $\Theta_k$, respectively, which are given as
\begin{align}\label{P}
\mathbf{O}_{B_p}=[\zeta_{1B_p},\zeta_{2B_p},\ldots,\zeta_{KB_p}],
\end{align}
\begin{align}\label{Q}
\mathbf{U}_{B_p}=[\eta_{1B_p},\eta_{2B_p},\ldots,\eta_{KB_p}],
\end{align}
and
\begin{align}\label{DTheta}
\mathbf{\Theta}=[\Theta^T_1,\Theta^T_2,\ldots,\Theta^T_K]^T.
\end{align}
Similarly, $\mathbf{O}_{E_q}$, $\mathbf{U}_{E_q}$ and $\Theta$ can be obtained. Then, the optimization problem is reduced as
\begin{subequations}\label{P10}
\begin{align}
&\mathop {{\max}}\limits_{\Theta_k}  ~\min\limits_{p,q} \{\log(1+\frac{|\mathbf{O}_{B_p}\mathbf{\Theta}|^2}{|\mathbf{U}_{B_p}\mathbf{\Theta}|^2+\sigma^2})\nonumber\\
&~~~~~~~~~~~~~-\log(1+\frac{|\mathbf{O}_{E_q}\mathbf{\Theta}|^2}{|\mathbf{U}_{E_q}\mathbf{\Theta}|^2+\sigma^2})\},\label{P10_1}\\
&~\mathrm{s}.\mathrm{t}.~~|\Theta_{k_m}|^2=1(\forall k=1,\ldots,K,\forall m=1,\ldots,M).\label{P10_2}
\end{align}
\end{subequations}
According to (\ref{P8}), problem (\ref{P10}) can be transformed as
\begin{subequations}\label{P11}
\begin{align}
&\mathop {{\max}}\limits_{\mathbf{v},\mathbf{w}}  ~\min\limits_{p,q} \{a_p-b_p-i_q+l_q\},\label{P11_1}\\
&~\mathrm{s}.\mathrm{t}.~~R_{\Theta_{m,m}}=1,\forall m=1,\ldots,KM,\label{P11_2}\\
&~~~~~~~\mathrm{Tr}(R_{O_{B_p}}R_{\Theta})+\mathrm{Tr}(R_{U_{B_p}}R_{\Theta})+\sigma^2\geq e^{a_p},\forall p,\label{P11_3}\\
&~~~~~~~\mathrm{Tr}(R_{U_{B_p}}R_{\Theta})+\sigma^2\leq e^{\bar{b_p}}(b_p-\bar{b_p}+1),\forall p,\label{P11_4}\\
&~~~~~~~\mathrm{Tr}(R_{O_{E_q}}R_{\Theta})+\mathrm{Tr}(R_{U_{E_q}}R_{\Theta})+\sigma^2\leq \nonumber\\
&~~~~~~~~~~~~~~~~~~~~~~~~~~~~~~~~~~~e^{\bar{i_q}}(i_q-\bar{i_q}+1),\forall q,\label{P11_5}\\
&~~~~~~~\mathrm{Tr}(R_{U_{E_q}}R_{\Theta})+\sigma^2\geq e^{l_q},\forall q,\label{P11_6}\\
&~~~~~~~R_v,R_w\succeq 0.\label{P11_7}
\end{align}
\end{subequations}
where $R_{O_{B_p}}=\mathbf{O}^H_{B_p}\mathbf{O}_{B_p}$, $R_{U_{B_p}}=\mathbf{U}^H_{B_p}\mathbf{U}_p$, $R_{O_{E_q}}=\mathbf{O}^H_{E_q}\mathbf{O}_{E_q}$, and $R_{U_q}=\mathbf{U}^H_{E_q}\mathbf{U}_{E_q}$ are known. $R_{\Theta}=\mathbf{\Theta}\mathbf{\Theta}^H$ satisfies that the diagonal element equals to one and rank$(R_{\Theta})=1$. Since the rank-one constraint is non-convex, we relax this constraint. Moreover, a set of exponential variables $a_p$, $b_p$, $i_q$ and $l_q$ are introduced which are substituted as
\begin{align}\label{evariable1}
&e^{a_p}=\mathrm{Tr}(R_{O_{B_p}}R_{\Theta})+\mathrm{Tr}(R_{U_{B_p}}R_{\Theta})+\sigma^2,\\
&e^{b_p}=\mathrm{Tr}(R_{U_{B_p}}R_{\Theta})+\sigma^2,\\
&e^{i_q}=\mathrm{Tr}(R_{O_{E_q}}R_{\Theta})+\mathrm{Tr}(R_{U_{E_q}}R_{\Theta})+\sigma^2,\\
&e^{l_q}=\mathrm{Tr}(R_{U_{E_q}}R_{\Theta})+\sigma^2,
\end{align}
and $\bar{b_p}$, $\bar{i_q}$ are the approximation point which are made around
\begin{align}
&\bar{b_p}=\ln\left(\mathrm{Tr}(R_{U_{B_p}}R_{\Theta})+\sigma^2 \right),\label{appr2_1}\\
&\bar{i_q}=\ln\left(\mathrm{Tr}(R_{O_{E_q}}R_{\Theta})+\mathrm{Tr}(R_{U_{E_q}}R_{\Theta})+\sigma^2 \right)\label{appr2_2}.
\end{align}
As a convex problem, (\ref{P11}) can be solved iteratively by the existing convex optimization solutions such as CVX. While the relaxed constraint may not lead to a rank-one solution, the Gaussian randomization can be similarly used as in problem (\ref{P10}) to obtain a feasible solution of (\ref{P11}) based on the higher-rank solution obtained by solving (\ref{P11}).

\begin{algorithm}[t]
\caption{Proposed MSR scheme}
\label{alg:A}
\begin{algorithmic}[1]
\STATE {Initialize $\mathbf{v}[0]$, $\mathbf{w}[0]$ and $\Theta[0]$ randomly that is feasible to (\ref{P4});}
\STATE {Set $n=0$;}
\STATE Set $m=0$;
\REPEAT
\STATE Substitute $\Theta[n]$ into (\ref{P8});
\REPEAT
\STATE {Substitute $\mathbf{v}[m]$ and $\mathbf{w}[m]$ into (\ref{appr1_1}) and (\ref{appr1_2}), respectively, yields $\bar{\nu_p}[m]$ and $\bar{\lambda_q}[m]$;}
\STATE Substitute $\bar{\nu_p}[m]$ and $\bar{\lambda_q}[m]$ into (\ref{P8}) yields the optimal solution ${\mathbf{R}_v}[m+1]$ and ${\mathbf{R}_w}[m+1]$;
\STATE Obtain $\mathbf{v}[m+1]$ and $\mathbf{w}[m+1]$ by Gaussian randomization;
\STATE {Update $m=m+1$;}
\UNTIL{Convergence}
\REPEAT
\STATE {Substitute $\mathbf{\Theta}[n]$ into (\ref{appr2_1}) and (\ref{appr2_2}), yields $\bar{b_p}[n]$ and $\bar{i_q}[n]$;}
\STATE Substitute $\bar{b_p}[n]$ and $\bar{i_q}[n]$ into (\ref{P11}) yields the optimal solution ${\mathbf{R}_{\Theta}}[n+1]$;
\STATE Obtain $\mathbf{\Theta}[n+1]$ by Gaussian randomization;
\STATE {Update $n=n+1$;}
\UNTIL{Convergence}
\UNTIL{Convergence}
\STATE {Output the final optimal solution $\mathbf{v}_{opt}=\mathbf{v}[m]$ , $\mathbf{w}_{opt}=\mathbf{w}[m]$ and $\Theta_{opt}=\Theta[n]$.}
\end{algorithmic}
\end{algorithm}
The detailed process of the MSR scheme is shown in Algorithm 2. First, we give initialized $\mathbf{v}[0]$, $\mathbf{w}[0]$ and $\mathbf{\Theta}[0]$, then the values of $\mathbf{v}[1]$, $\mathbf{w}[1]$ are obtained with $\mathbf{\Theta}[0]$ by solving problem (\ref{P8}). Next, we obtain $\mathbf{\Theta}[1]$ with $\mathbf{v}[1]$, $\mathbf{w}[1]$ by solving problem (\ref{P11}). Thus, the optimal solutions of $\mathbf{v}$, $\mathbf{w}$ and $\mathbf{\Theta}$ can be obtained within a finite number of iterations.

The convergence is analyzed as follows. Algorithm 2 consists of twice CVX optimizations and once iteration operation. In the two CVX algorithms, since the objective function is concave that has a maximum value, and all constraints are convex, the CVX algorithms converge. From step 6 to step 11, we can obtain $\mathbf{v}[m+1]$ and $\mathbf{w}[m+1]$ which satisfy $F(\mathbf{v}[m+1]),\mathbf{w}[m+1],\mathbf{\Theta}[n]\geq F(\mathbf{v}[m]),\mathbf{w}[m],\mathbf{\Theta}[n]$, where $F$ denotes the objective function. From step 12 to step 17, $\mathbf{\Theta}[n]$ can be obtained which satisfies $F(\mathbf{v}[m+1]),\mathbf{w}[m+1],\mathbf{\Theta}[n+1]\geq F(\mathbf{v}[m+1]),\mathbf{w}[m+],\mathbf{\Theta}[n]$. Therefore, Algorithm 2 converges.

Next, we study the complexity of the proposed algorithm. According to \cite{book2012}, the per-iteration complexity of the proposed high performance scheme can be approximately calculated as $O(T_1N^{3.5}+T_2(MK)^{3.5}+((K^2+2K)M^2+(K+1)N^2+KMN)(P+Q))$, where $T_1$ and $T_2$ denote the inner iterations times from step 6 to step 11 and from step 12 to step 17 in Algorithm 2, respectively.

\subsection{A low-complexity MSLNR scheme}
Since the complexity of the MMSR scheme proposed in Section III-A is high in practice, we propose a low-complexity MSLNR scheme in this subsection which is more practical in realistic scenarios. Based on the MSLNR scheme, the beamforming vector on Alice and the phase shifting on RIS are designed to preserve the power of confidential message as possible in the desired users by minimizing the leakage of confidential message to the eavesdroppers (undesired users). Meanwhile, the AN vector is designed to minimize its impact on desired users, i.e. designed on the null-space of the desired users. The benefit of this scheme is the significantly reduced complexity compared to the MSR scheme.

First, we give the optimal problem of MSLNR scheme as
\begin{subequations}\label{PSLNR}
\begin{align}
&\mathop {{\max}}\limits_{\mathbf{v},\mathbf{w},\Theta}  ~\mathrm{SLNR}\label{PSLNR_1}\\
&~\mathrm{s}.\mathrm{t}.~~\|\mathbf{v}\|^2=1,\label{PSLNR_2}\\
&~~~~~~~\|\mathbf{w}\|^2=1  \label{PSLNR_3}\\
&~~~~~~~(\ref{Theta}),\forall k=1,2,\ldots,K.\label{PSLNR_4}
\end{align}
\end{subequations}

According to the definition of SLNR, the expression can be given by
\begin{align}\label{SLNR}
\mathrm{SLNR}=\frac{|\sqrt{\alpha}\mathbf{v}^T\mathbb{H}_{\Theta B}|^2}{|\sqrt{\alpha}\mathbf{v}^T\mathbb{H}_{\Theta E}|^2+\sigma^2}
\end{align}
where
\begin{align}\label{HHb}
\mathbb{H}_{\Theta B}=[&\sum \limits_{k=1}^K \Upsilon_{k1} \Omega_k \mathbf{H}^H_{I_kB_1}(\mathbf{f})\mathbf{\Theta}_k,\ldots,\nonumber\\
&\sum \limits_{k=1}^K \Upsilon_{kP} \Omega_k \mathbf{H}^H_{I_kB_P}(\mathbf{f})\mathbf{\Theta}_k]
\end{align}
and
\begin{align}\label{HHe}
\mathbb{H}_{\Theta E}=[&\sum \limits_{k=1}^K \Upsilon_{k1} \Omega_k \mathbf{H}^H_{I_kB_1}(\mathbf{f})\mathbf{\Theta}_k,\ldots,\nonumber\\
&\sum \limits_{k=1}^K \Upsilon_{kQ} \Omega_k \mathbf{H}^H_{I_kB_Q}(\mathbf{f})\mathbf{\Theta}_k].
\end{align}
In this case, with a given $\mathbf{\Theta}$, when $\mathbb{H}_{\Theta B}$ and $\mathbb{H}_{\Theta E}$ are known, (\ref{SLNR}) can be rewritten as
\begin{align}\label{SLNR1}
\mathrm{SLNR}=\frac{\alpha(\mathbf{v^*})^H\mathbb{H}_{\Theta B_p}\mathbb{H}^H_{\Theta B_p}\mathbf{v^*}}{\alpha(\mathbf{v^*})^H\mathbb{H}_{\Theta E_p}\mathbb{H}^H_{\Theta E_p}\mathbf{v^*}+\sigma^2}.
\end{align}
Thus, according to the generalized Rayleigh-Ritz theorem \cite{Horn1985Matrix}, the optimal solution of $\mathbf{v^*}$ is given by the largest eigen-value of the matrix
\begin{align}\label{eigen}
[\mathbb{H}_{\Theta E_p}\mathbb{H}^H_{\Theta E_p}+\frac{\sigma^2}{\alpha}\mathbf{I}_N]^{-1}\mathbb{H}_{\Theta B_p}\mathbb{H}^H_{\Theta B_p}.
\end{align}
Then, with the obtained $\mathbf{v}$, the SLNR can be expressed as
\begin{align}\label{SLNR2}
\mathrm{SLNR}=\frac{\sum^P \limits_{p=1}|\mathbf{O}_{B_p}\mathbf{\Theta}|^2}{\sum^Q \limits_{q=1}|\mathbf{O}_{E_q}\mathbf{\Theta}|^2+\sigma^2}.
\end{align}
Next, after we give an initial value of $\Theta[0]$, the optimal $\varphi_{k,m}$ can be obtained by the gradient descent method when $\varphi_{k',m'}(k' \neq k,m'\neq m)$ is fixed. Herein, $\varphi_{k,m}=\varphi_{k_{mc,mr}}$, where $m=(mc-1)Mr+mr$. The specific derivation of $\varphi_{k,m}$ is given in Appendix A.

Accordingly, $\Theta[1]$ is obtained, after a finite number of iterations, the optimal solution of $\mathbf{v}$ and $\Theta$ can be obtained.
\begin{algorithm}[t]
\caption{Proposed MSLNR scheme}
\label{alg:C}
\begin{algorithmic}[1]
\STATE {Initialize $\mathbf{v}[0]$ and $\mathbf{\Theta}[0]$ randomly that is feasible to (\ref{Theta}), and $\mathbf{v}^H[0]\mathbf{v}[0]=1$, respectively;}
\STATE {Set $n=0$;}
\REPEAT
\STATE Set $m=0$;
\REPEAT
\STATE {Substitute $\mathbf{v}[n]$ and $\mathbf{\Theta}[m]$ into (\ref{SLNR}) yields $\mathbf{\Theta}[m+1]$ according to Appendix A;}
\STATE {Update $m=m+1$;}
\UNTIL{Convergence}
\STATE Substitute $\mathbf{\Theta}[m]$ into (\ref{HHb}),(\ref{HHe}) and (\ref{eigen}) yields the optimal solution $\mathbf{v}[n+1]$;
\STATE Update $n=n+1$;
\UNTIL{Convergence}
\STATE {Output the final optimal solution $\mathbf{v}_{opt}=\mathbf{v}[n]$ and $\mathbf{\Theta}_{opt}=\mathbf{\Theta}_{opt}[m]$.}
\end{algorithmic}
\end{algorithm}
The complete iterative procedure for solving problem (\ref{PSLNR}) is summarized in Algorithm \ref{alg:C}.

In Algorithm \ref{alg:C}, we first initialize  $\mathbf{v}[0]$ and $\mathbf{\Theta}[0]$ which satisfy their constraints. Then, we yield $\mathbf{\Theta}[1]$ with the initial value of $\mathbf{\Theta}[0]$ and $\mathbf{v}[0]$ according to Appendix A. Next, with $\mathbf{\Theta}[1]$, we can obtain $\mathbf{v}[1]$ by the generalized Rayleigh-Ritz theorem. Thus, after a number of iterations, the algorithm converges, and the optimal solutions of $\mathbf{\Theta}$ and $\mathbf{v}$ are obtained. At last, AN vector is designed to minimize the impact on desired user by designing it on the null-space of the desired users. The optimization problem is given by
\begin{align}\label{P12}
&\min \limits_{\mathbf{w}} ~|\sqrt{1-\alpha}\mathbf{w}^T\mathbb{H}_{\Theta B}|^2\nonumber\\
&s.t ~~~~~\mathbf{w}^T\mathbf{w}=1.
\end{align}
Let $\Gamma_p =\sum \limits_{k=1}^K \Upsilon_{kp} \Omega_k \mathbf{H}^H_{I_kB_p}(\mathbf{f})\mathbf{\Theta}_k$, $\Gamma \in \mathbb{C}(N\times 1)$, then $\mathbb{H}_{\Theta B}$ is reduced as
\begin{align}\label{HHb1}
\mathbb{H}_{\Theta B}=[\Gamma_1,\Gamma_2,\ldots,\Gamma_P].
\end{align}
make a reasonable assumption of $N\geq P$, which means that the number of transmit antennas at Alice is larger than the number of desired users. To project the AN on the null space of desired users, the following orthogonal projector is constructed as
\begin{align}\label{wzj}
\mathbf{w}^T=\frac{\mathbf{z}(\mathbf{I}_N-\mathbb{H}_{\Theta B}[\mathbb{H}^H_{\Theta B}\mathbb{H}_{\Theta B}]^{-1}\mathbb{H}^H_{\Theta B})}{|\mathbf{z}(\mathbf{I}_N-\mathbb{H}_{\Theta B}[\mathbb{H}^H_{\Theta B}\mathbb{H}_{\Theta B}]^{-1}\mathbb{H}^H_{\Theta B})|}.
\end{align}
In this case, we have
\begin{align}\label{wH}
\mathbf{w}^T\mathbb{H}_{\Theta B}=\frac{\mathbf{z}(\mathbb{H}_{\Theta B}-\mathbb{H}_{\Theta B})}{|\mathbf{z}(\mathbf{I}_N-\mathbb{H}_{\Theta B}[\mathbb{H}^H_{\Theta B}\mathbb{H}_{\Theta B}]^{-1}\mathbb{H}^H_{\Theta B})|}=\mathbf{0}_{1\times P}.
\end{align}
This result means that the received AN power of each desired user is zero, thus AN only distorts the received signal at undesired users.
In conclusion, the optimal solution of $\mathbf{v}$, $\mathbf{w}$, and $\mathbf{\Theta_k},\forall k=1,\ldots,K$ are obtained by the proposed MSLNR scheme.

Finally, we analyze the convergence and complexity of Algorithm \ref{alg:C}. First of all, it is clear to observe that $\mathrm{SLNR}(\mathbf{v}[n+1],\mathbf{\Theta}[m+1])\geq \mathrm{SLNR}(\mathbf{v}[n+1],\mathbf{\Theta}[m])\geq \mathrm{SLNR}(\mathbf{v}[n],\mathbf{\Theta}[m])$, thus the algorithm converges. The per-iteration complexity of MSLNR scheme can be approximately calculated as $O(3N^3+P^3+(MN(N+K+1)+N^2+N+TKM)(P+Q)+NP(N+P))$, where $T$ is the times of inner iterations from step 4 to step 8 in Algorithm \ref{alg:C}.

\section{simulation results and analysis}
In this section, we evaluate the performance of our proposed schemes by numerical simulations. The default system parameters are chosen as shown in Table \ref{tab}. By default, we assume $\sigma^2_B=\sigma^2_E$.
\begin{table}[t]
\centering
\caption{SIMULATION PARAMETERS SETTING}\label{tab}
 \begin{tabular}{|c|c|}
\hline
Parameter&Value\\
\hline
The number of transmitter antennas ($N_r\times N_c$)&$4\times 4$ \\
\hline
The number of RIS elements ($M_r\times M_c$)&$4\times 4$ \\
\hline
The number of desired users (P)&2\\
\hline
The number of eavesdroppers (Q)&2\\
\hline
The number of RIS (K)&2\\
\hline
The antenna spacing of Alice ($d_A$)&$c/2f_c$\\
\hline
The element spacing of RIS ($d_I$)&$c/2f_c$\\
\hline
Total signal bandwidth (B)&5MHz\\
\hline
Total transmit power ($P$)&1W \\
\hline
Power allocation factors for MSLNR scheme&0.9\\
\hline
\tabincell{c}{Desired users' positions \\ $({(x_{B_1},y_{B_1},z_{B_1}),(x_{B_2},y_{B_2},z_{B_2})})$}&\tabincell{c}{$(100m,50m,0)$,\\$(200m,150m,0)$}\\
\hline
\tabincell{c}{Eavesdroppers' positions \\
$({(x_{E_1},y_{E_1},z_{E_1}),(x_{E_2},y_{E_2},z_{E_2})})$}&\tabincell{c}{$(150m,0,0)$,\\$(200m,50m,0)$}\\
\hline
\tabincell{c}{RISs' positions \\
$({(x_{I_1},y_{I_1},z_{I_1}),(x_{I_2},y_{I_2},z_{I_2})})$}&\tabincell{c}{$(50m,150m,50m)$,\\$(100m,200m,30m)$}\\
\hline
Central carrier frequency ($f_c$)&3GHz\\
\hline
Subcarriers number (N)&1024\\
\hline
\end{tabular}
\end{table}

\begin{figure}[t]
\centering
\includegraphics[width=0.50\textwidth]{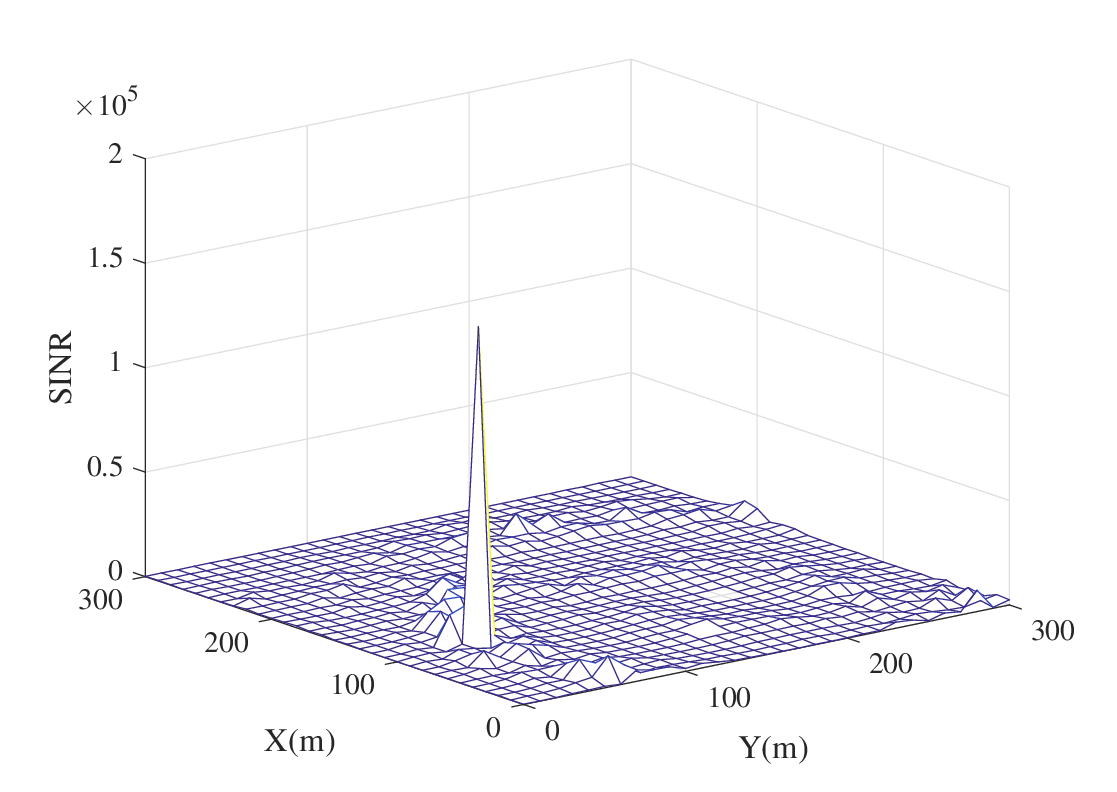}\\
\caption{3-D surface of SINR versus coordinate of proposed MSINR scheme.}\label{SINR}
\end{figure}

\begin{figure}[t]
\centering
\includegraphics[width=0.50\textwidth]{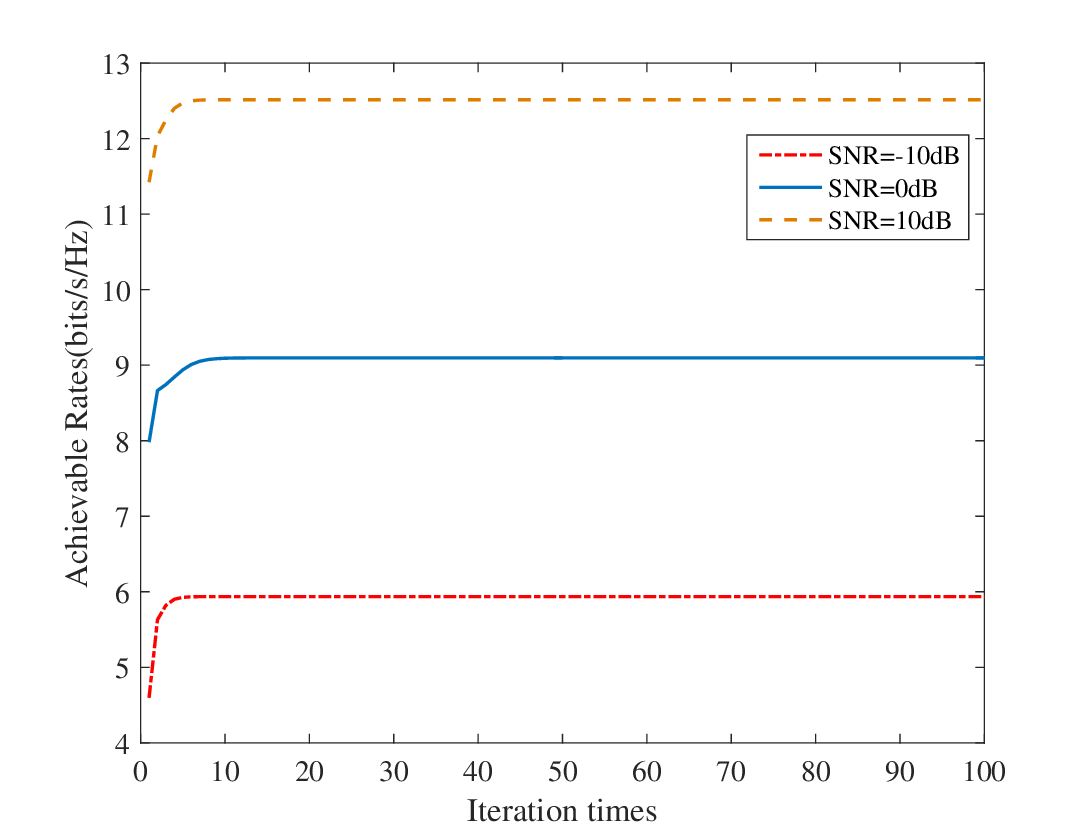}\\
\caption{Achievable rate versus iteration times of proposed MSINR scheme.}\label{iteration}
\end{figure}
Fig. \ref{SINR} illustrates the 3-D performance surface of SINR versus the location coordinate of the proposed MSINR method, since with the assumption that the users are all on the ground i.e. the Z axis coordinate is $0$, and the SINR versus the X and Y axes coordinate is drawn. It can be observed that, in the single-user scenario, proposed MSINR scheme has achieved SPWT since there is only one high signal energy peak of confidential messages formed in the desired position (in this scenario, the single user set on $(100m,50m,0)$ is considered), while outside the main peak, the SINR is far lower than that of the desired user. The average SINR outside the main peak is only less than one tenth of that in the desired user. Moreover, the SINR versus the iteration times with random initial value of $\mathbf{v}$ and $\Theta$ is illustrated in Fig. \ref{iteration}. It is seen from the figure that the MSINR algorithm converges within $10$ iterations. According to Section III, each iteration has an analytical solution, thus the MSINR scheme based on single user scenario has a extremely low complexity for practical applications.

\begin{figure}[t]
\centering
\includegraphics[width=0.50\textwidth]{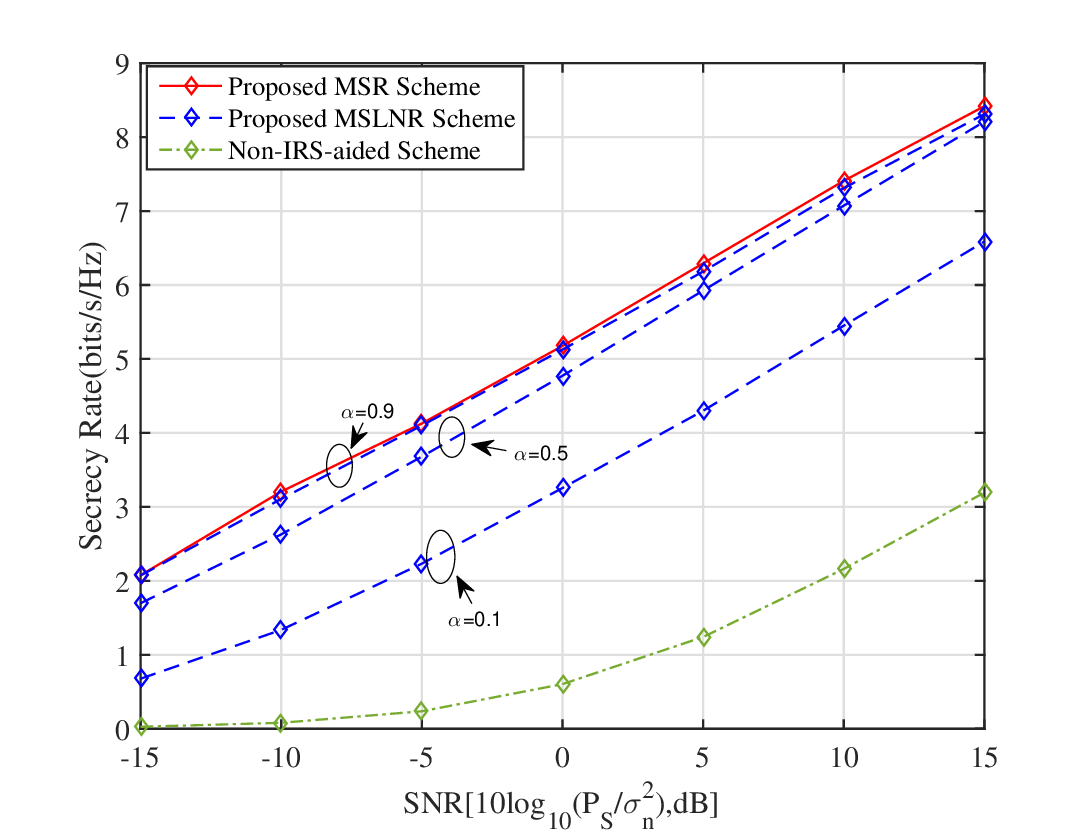}\\
\caption{Curves of SR versus SNR for proposed schemes in multi-user scenario.}\label{SR}
\end{figure}

Fig. \ref{SR} draws the curves of secrecy rates versus SNR based on proposed MSR scheme, proposed MSLNR scheme and conventional scheme without RIS. Besides, the conventional scheme is based on \cite{Shu2018SPWT}, which has the available direct path channel. The simulation result show that, the MSR scheme has the best secrecy rate performance, and the MSLNR scheme has a slightly lower secrecy performance than the MSR scheme. Moreover, even if the conventional scheme has the direct path, its secrecy rate performance still has a obvious gap compared to the former two schemes. This result show that our proposed multi-RIS schemes have the spectrum efficiency advantage. Additionally, the value of power allocation factor $\alpha$ is also related to the SR performance of the MSLNR scheme, which is illustrated in Fig.\ref{SR}. With three different $\alpha$ values of with $0.9$, $0.5$ and $0.1$, the SR performances are significantly different. The SR performance improves as the value of $\alpha$ increases. Since the value of $\alpha$ is intimately related to energy efficiency, how to choose the value of $\alpha$ should consider a balance among energy efficiency, security and performance.
\begin{figure}[t]
\centering
\includegraphics[width=0.50\textwidth]{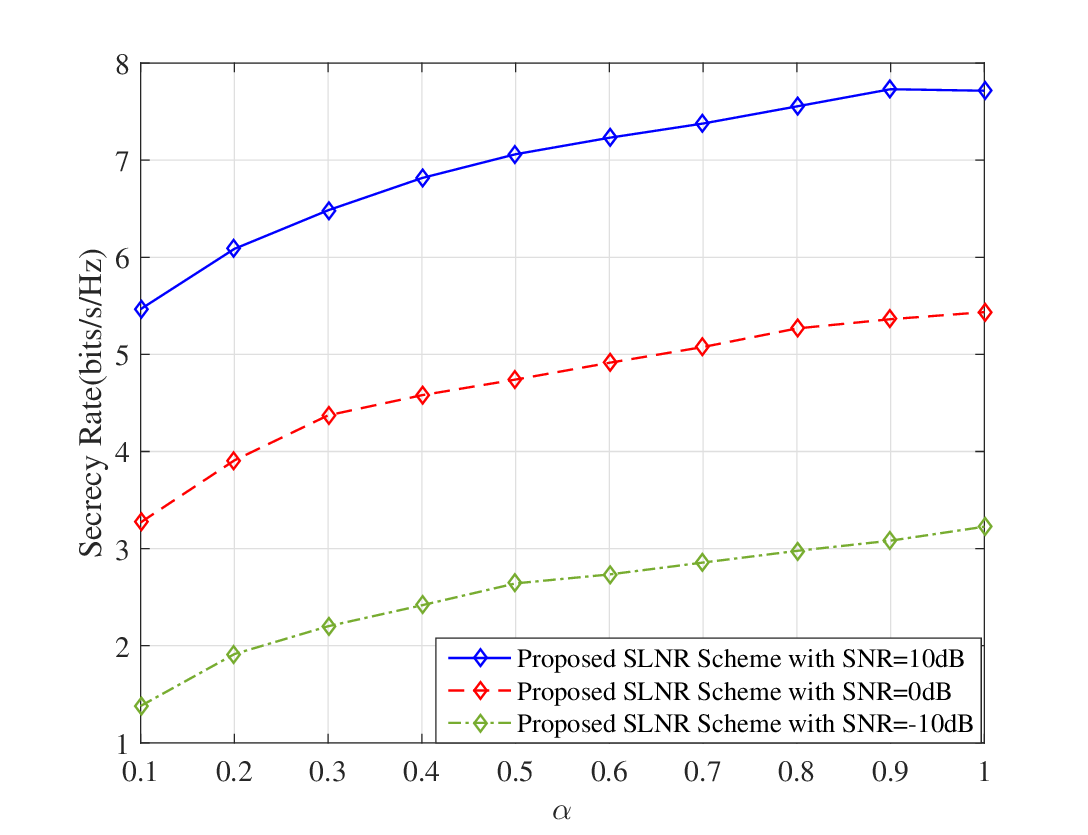}\\
\caption{Curves of SR versus power allocation factor $\alpha$ for proposed MSLNR scheme.}\label{Alpha}
\end{figure}

Based on the above discussion, we make an investigation concerning the impact of $\alpha$ on the SR performance in Fig. \ref{Alpha}. Fig. \ref{Alpha} draws the curves of SR versus $\alpha$ in three different SNR scenarios: (1)SNR=-10dB, (2)SNR=0dB and (3)SNR=10dB. It can be seen that, in the low SNR region, the impact of AN is weak due to a large channel noise, thus the SR curve increases with increasing $\alpha$ and the value of $\alpha$ should be as large as possible. However, as the SNR increases, the AN plays a key role in SR performance, the curve is a concave function of $\alpha$, and there exists the optimal value of $\alpha$.

\begin{figure}[t]
\centering
\includegraphics[width=0.50\textwidth]{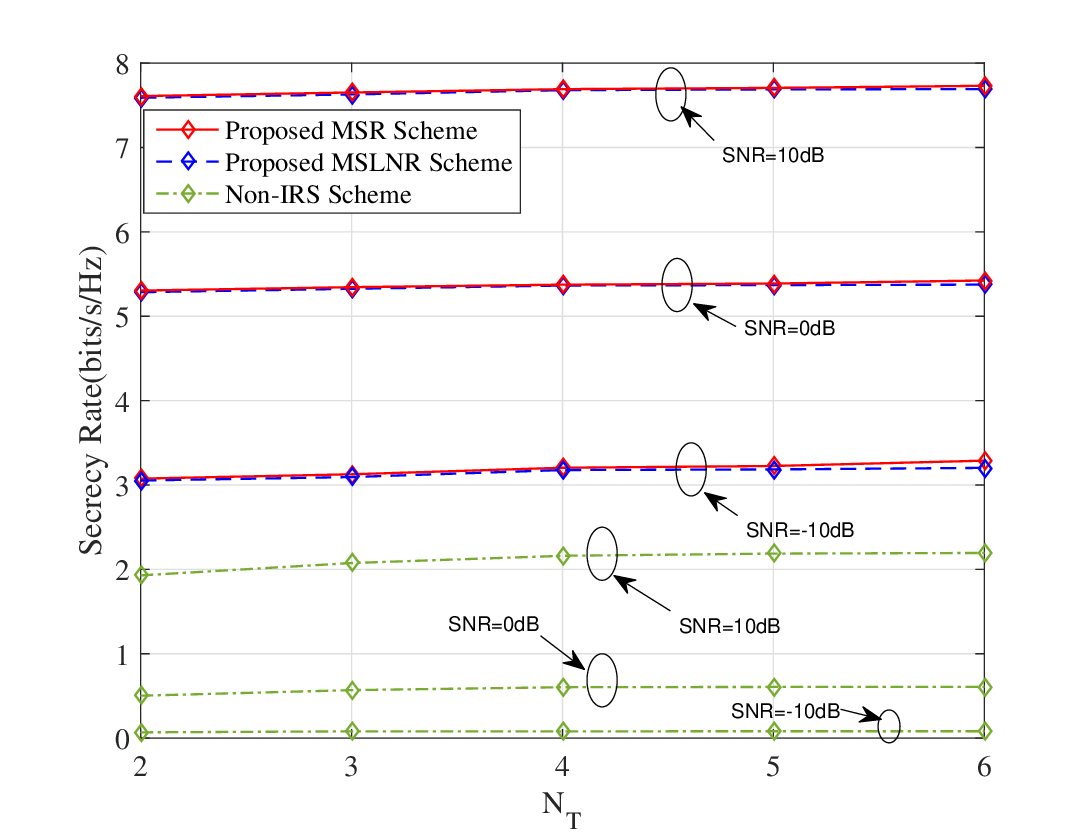}\\
\caption{Curves of SR versus $N_T$ for proposed schemes in multi-user scenario.}\label{NT}
\end{figure}
Fig. \ref{NT} shows the impact of transmit antenna number on the SR performance. Since the transmitter is a $N_r\times N_c$ rectangular antennas array, for convenience we set $N_r=N_c=N_T$ and the total number of antenna is $N_T^2$, and thus Fig. \ref{NT} shows the secrecy rates versus $N_T$. Observing Fig. \ref{NT}, with the increasing of the antenna number, the improvement of secrecy rate is not obvious. This is due to the fact that transmitting power and the number of reflection path are both fixed, thus increasing the number of transmit antennas can not improve the received confidential signal power significantly.

\begin{figure}[t]
\centering
\includegraphics[width=0.50\textwidth]{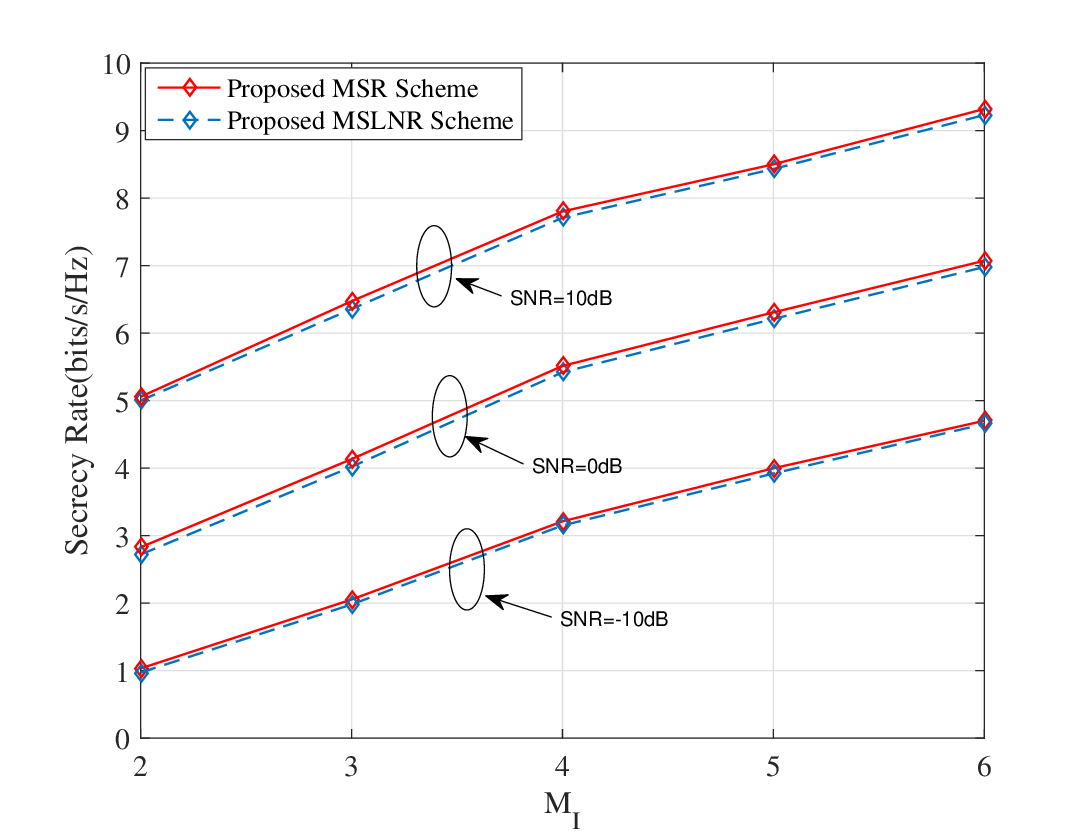}\\
\caption{Curves of SR versus $M_I$ for proposed schemes in multi-user scenario.}\label{MI}
\end{figure}
In Fig. \ref{MI}, the impact of RIS element number on the SR performance is evaluated. Similarly, we set $M_r=M_c=M_I$ and the total number of RIS element is $M_I^2$. Note that the performance of non-RIS scheme is independent on $M_I$, thus the curves of non-RIS scheme are neglected. The result is different from that in Fig. \ref{NT}. It can be seen that, with the increase of the RIS element number, the secrecy rate is improved significantly. The reason for this result is that, each RIS element reflects confidential message transmitted from Alice. Increasing the number of RIS elements can be regarded as increasing the number of reflection path. Thus, with the optimal designs of beamforming vector phase shifting, the more the reflection paths, the more confidential message energy can be received by desired users, thus the SR increases.

In summary, the proposed MSINR scheme can transmit confidential signal securely and precisely, and only a single energy peak yields in the desired location. Moreover, the power allocation factor will affect the secrecy rate, and there exists an optimal value. Then, the proposed MSR scheme has the optimal secrecy performance, the proposed MSLNR scheme has the low-complexity and a close secrecy performance to the MSR scheme. Moreover, the number of Alice's antennas has slight impact on secrecy rate but the number of RIS elements affecting significantly in secrecy rate, namely, with the elements number of RIS increases, the secrecy rate increases.

%
%
%

\section{Conclusion}
In this paper, multi-RISs-aided schemes have been proposed to achieve SPWT based on the scenario that the direct path between transmitter and receivers are unavailable. First, a MSINR scheme was proposed based on a single-user scenario while there exist potential eavesdroppers and their position knowledge is unavailable. With the proposed MSINR scheme, the desired user can achieve a maximum received confidential signal energy. Out of the desired user's position, confidential signal is protected from eavesdropping by beamforming, phase shifting and AN. Only weak confidential signal energies are conserved in these locations. Next, a high performance MSR scheme and a low-complexity MSLNR scheme have been proposed in multi-user scenario. Using these two schemes, we obtained the following interesting results: (1) the proposed MSR scheme has the superior SR performance, and proposed MSLNR scheme has a slightly lower performance than that of the MSR scheme but has a significantly lower complexity. The two schemes perform much better than non-RIS scheme. (2) The SR performances of the proposed two schemes are related to the transmit antenna number and the RIS element number, and the SR performance improvement of RIS element number is more obvious. (3) The power allocation factor can impact the SR performance of proposed MSLNR scheme. As the SNR increases, AN becomes more and more important. Due to the high security and low complexity of RIS, the proposed schemes can be potentially applied in the future scenarios including UAV communications, automobile communications and so on.

\begin{appendices}

\section{DERIVATION OF $\varphi_{m}$}
By substituting Eqs. (\ref{P}) and (\ref{DTheta}) into $\sum^P \limits_{p=1}|\mathbf{O_{B_p}}\mathbf{\Theta}|^2$, we have
\begin{align}\label{hbvavd}
\mathbf{O}_{B_p}\mathbf{\Theta}&= \sum^K \limits_{k=1} \sum^M \limits_{m=1} \mathbf{O}_{B_{p_{k,m}}}{e^{j{\varphi_{k,m}}}} \nonumber\\
 &= {e^{j{\varphi _{k,m}}}}{A_{{B_{p_{k,m}}}}} + {B_{{B_{p_{k,m}}}}},\forall p=1,\ldots,P
\end{align}
where
\begin{align}\label{ABm}
&{A_{{B_{p_{k,m}}}}}={  O_{B_{p_{k,m}}}},\\
&{B_{{B_{p_{k,m}}}}}=\sum^{K,M} \limits_{(k',m')\neq (k,m)} \mathbf{O}_{B_{p_{k',m'}}}{e^{j{\varphi_{k',m'}}}}.
\end{align}

Then, we have
\begin{align}\label{hbvavd2}
\sum^P \limits_{p=1}|\mathbf{O}_{B_p}\mathbf{\Theta}|^2&=\sum^P \limits_{p=1}({A_{{B_{p_{k,m}}}}}e^{j{\varphi_{k,m}}} + {B_{{B_{p_{k,m}}}}})^H\cdot\nonumber\\
&~~~~~~~~({{A_{{B_{p_{k,m}}}}}e^{j{\varphi_{k,m}}}} + {B_{{B_{p_{k,m}}}}})\nonumber\\
&=\sum^P \limits_{p=1}(A_{{B_{p_{k,m}}}}^HA_{{B_{p_{k,m}}}}+B_{{B_{p_{k,m}}}}^HB_{{B_{p_{k,m}}}})\nonumber\\
&~+2(R_{B_{{k,m}}}\cos \varphi_{k,m}-I_{B_{{k,m}}}\sin \varphi_{k,m}),
\end{align}
with
\begin{align}\label{RBm1}
&R_{B_{k,m}}=\sum^P \limits_{p=1}\mathrm{Re}(A_{{B_{p_{k,m}}}}B_{{B_{p_{k,m}}}}^H),
\end{align}
\begin{align}\label{RBm2}
&I_{B_{k,m}}=\sum^P \limits_{p=1}\mathrm{Im}(A_{{B_{p_{k,m}}}}B_{{B_{p_{k,m}}}}^H).
\end{align}
where $\mathrm{Re}(\cdot)$ and $\mathrm{Im}(\cdot)$ denote the real part and the imaginary part, respectively.

Similarly, we can obtain $|\mathbf{O}_{E_q}\mathbf{\Theta}|^2$, $A_{{E_{q_{k,m}}}}$, $B_{{E_{q_{k,m}}}}$, $R_{E_{{k,m}}}$ and $I_{E_{{k,m}}}$ where $q=1,\ldots,Q$. Thus, the SLNR can be expressed as
\begin{align}\label{SLNR2}
\mathrm{SLNR}=\frac {X_{B_{k,m}}+2 R_{B_{k,m}} \cos \varphi_{k,m} - 2 I_{B_{k,m}}\sin \varphi_{k,m}} {X_{E_{k,m}}+2 R_{E_{k,m}} \cos \varphi_{k,m} - 2 I_{E_{k,m}}\sin \varphi_{k,m}}
\end{align}
where
\begin{align}\label{XB}
&X_{B_{k,m}}=\sum^P \limits_{p=1}(A_{{B_{k,m}}}^HA_{{B_{k,m}}}+B_{{B_{k,m}}}^HB_{{B_{k,m}}}),\\
&X_{E_{k,m}}=\sum^P \limits_{p=1}(A_{{E_{k,m}}}^HA_{{E_{k,m}}}+B_{{E_{k,m}}}^HB_{{E_{k,m}}})+\sigma^2.
\end{align}

Therefore, in order to obtain the optimal $\varphi_m$, we compute the derivation of $\mathrm{SLNR}_B$ with respect to $\varphi_m$,
\begin{align}\label{deri}
&\frac{\partial(\mathrm{SLNR})}{\partial (\varphi_{k,m})}=\nonumber\\
&\frac {2y_1\sin \varphi_{k,m}+2y_2\cos \varphi_{k,m}+4R_{B_{k,m}}I_{E_{k,m}}-4R_{E_{k,m}}I_{B_{k,m}}} {[X_{E_{k,m}}+2 Re_{E_{k,m}} \cos \varphi_{k,m} - 2 I_{E_{k,m}}\sin \varphi_{k,m}]^2},
\end{align}
where
\begin{align}\label{y1}
&y_1=X_{B_{k,m}}R_{E_{k,m}}-X_{E_{k,m}}R_{B_{k,m}},
\end{align}
\begin{align}\label{y2}
&y_2=X_{B_{k,m}}I_{E_{k,m}}-X_{E_{k,m}}I_{B_{k,m}}.
\end{align}
Let $\frac{\partial(\mathrm{SLNR})}{\partial (\varphi_{k,m})}=0$, we obtain
\begin{align}\label{phim}
\varphi_{k,m}=\arcsin \frac{2R_{E_{k,m}}I_{B_{k,m}}-2R_{B_{k,m}}I_{E_{k,m}}}{\sqrt{y_1^2+y_2^2}}-\varphi_y,
\end{align}
where $\varphi_y$ satisfies $\sin \varphi_y=\frac{y_2}{\sqrt{y_1^2+y_2^2}}$ and $\cos \varphi_y=\frac{y_1}{\sqrt{y_1^2+y_2^2}}$.

\noindent This complete the derivation of $\varphi_{k,m}$.

\end{appendices}

\ifCLASSOPTIONcaptionsoff
  \newpage
\fi
\bibliographystyle{IEEEtran}

\bibliography{IEEEfull,reference}

\begin{thebibliography}{10}
\providecommand{\url}[1]{#1}
\csname url@samestyle\endcsname
\providecommand{\newblock}{\relax}
\providecommand{\bibinfo}[2]{#2}
\providecommand{\BIBentrySTDinterwordspacing}{\spaceskip=0pt\relax}
\providecommand{\BIBentryALTinterwordstretchfactor}{4}
\providecommand{\BIBentryALTinterwordspacing}{\spaceskip=\fontdimen2\font plus
\BIBentryALTinterwordstretchfactor\fontdimen3\font minus
  \fontdimen4\font\relax}
\providecommand{\BIBforeignlanguage}[2]{{%
\expandafter\ifx\csname l@#1\endcsname\relax
\typeout{** WARNING: IEEEtran.bst: No hyphenation pattern has been}%
\typeout{** loaded for the language `#1'. Using the pattern for}%
\typeout{** the default language instead.}%
\else
\language=\csname l@#1\endcsname
\fi
#2}}
\providecommand{\BIBdecl}{\relax}
\BIBdecl

\bibitem{Babakhani2008}
A.~{Babakhani}, D.~{Rutledge}, and A.~{Hajimiri}, ``Transmitter architectures
  based on nearfield direct antenna modulation,'' \emph{IEEE J.Solid-State
  Circuits}, vol.~43, no.~12, pp. 2674--2692, Dec. 2008.

\bibitem{Babakhani2009}
------, ``Near-field direct antenna modulation,'' \emph{IEEE Microw.Mag},
  vol.~10, no.~1, pp. 36--46, Feb. 2009.

\bibitem{Daly2009Directional}
M.~P. Daly and J.~T. Bernhard, ``Directional modulation technique for phased
  arrays,'' \emph{IEEE Trans. Antennas Propagat.}, vol.~57, no.~9, pp.
  2633--2640, Sep. 2009.

\bibitem{Daly2010Directional}
M.~P. Daly, E.L.Daly, and J.~T. Bernhard, ``Demonstration of directional
  modulation using a phased array,,'' \emph{IEEE Trans. Antennas Propagat.},
  vol.~58, no.~5, pp. 1545--1550, May 2010.

\bibitem{Wang2012Distributed}
H.~Wang, Q.~Yin, and X.~Xia, ``Distributed beamforming for physical-layer
  security of two-way relay networks,'' \emph{IEEE Trans. Signal Process.},
  vol.~60, no.~7, pp. 3532--3545, Jul. 2012.

\bibitem{ChenX2017}
X.~Chen, D.~W.~K. Ng, W.~H. Gerstacker, and H.~H. Chen, ``A survey on
  multiple-antenna techniques for physical layer security,'' \emph{IEEE Commun.
  Surveys Tuts.}, vol.~19, no.~2, pp. 1027--1053, Nov. 2017.

\bibitem{Zou2016Relay}
Y.~Zou, J.~Zhu, X.~Li, and L.~Hanzo, ``Relay selection for wireless
  communications against eavesdropping: A security-reliability trade-off
  perspective,'' \emph{IEEE Network}, vol.~30, no.~5, pp. 74--79, Oct. 2016.

\bibitem{JZhao2019}
J.~{Zhao}, Q.~{Li}, Y.~{Gong}, and K.~{Zhang}, ``Computation offloading and
  resource allocation for cloud assisted mobile edge computing in vehicular
  networks,'' \emph{IEEE Trans. Veh. Technol.}, vol.~68, no.~8, pp. 7944--7956,
  Aug 2019.

\bibitem{Guo2017Exploiting}
J.~Guo, N.~Zhao, R.~Yu, X.~Liu, and V.~Leung, ``Exploiting adversarial jamming
  signals for energy harvesting in interference networks,'' \emph{IEEE Trans.
  Wirel. Commun.}, vol.~16, no.~2, pp. 1267--1280, Feb. 2017.

\bibitem{zhouTWC2022}
X.~Zhou, S.~Yan, Q.~Wu, F.~Shu, and D.~W.~K. Ng, ``Intelligent reflecting
  surface ({IRS})-aided covert wireless communications with delay constraint,''
  \emph{IEEE Trans. Wireless Commun.}, vol.~21, no.~1, pp. 532--547, Jan. 2022.

\bibitem{Wu2017Secure}
Y.~Wu, J.~B. Wang, J.~Wang, R.~Schober, and C.~Xiao, ``Secure transmission with
  large numbers of antennas and finite alphabet inputs,'' \emph{IEEE Trans.
  Commun.}, vol.~65, no.~8, pp. 3614--3628, Aug. 2017.

\bibitem{Ni2019}
S.~{Ni}, J.~{Zhao}, H.~H. {Yang}, and Y.~{Gong}, ``Enhancing downlink
  transmission in mimo hetnet with wireless backhaul,'' \emph{IEEE Trans. Veh.
  Technol.}, vol.~68, no.~7, pp. 6817--6832, July 2019.

\bibitem{Yan2016Artificial}
S.~Yan, X.~Zhou, N.~Yang, B.~He, and T.~Abhayapala, ``Artificial-noise-aided
  secure transmission in wiretap channels with transmitter-side correlation,''
  \emph{IEEE Trans. Wirel. Commun.}, vol.~15, no.~12, pp. 8286 -- 8297, Dec.
  2016.

\bibitem{Zhao2016Anti}
N.~Zhao, F.~R. Yu, M.~Li, and V.~C.~M. Leung, ``Anti-eavesdropping schemes for
  interference alignment ({IA})-based wireless networks,'' \emph{IEEE Trans.
  Wireless Commun.}, vol.~15, no.~8, pp. 5719--5732, May. 2016.

\bibitem{Hu2016}
J.~Hu, F.~Shu, and J.~Li, ``Robust synthesis method for secure directional
  modulation with imperfect direction angle,'' \emph{IEEE Commun.Lett},
  vol.~20, no.~6, pp. 1084 -- 1087, Jun. 2016.

\bibitem{Shu2016DM}
F.~{Shu}, X.~{Wu}, J.~{Li}, R.~{Chen}, and B.~{Vucetic}, ``Robust synthesis
  scheme for secure multi-beam directional modulation in broadcasting
  systems,'' \emph{IEEE Access}, vol.~4, pp. 6614--6623, 2016.

\bibitem{2019Optimal}
L.~U. Zaoyu, L.~Sun, S.~Zhang, X.~Zhou, J.~Lin, W.~Cai, J.~Wang, L.~U. Jinhui,
  and F.~Shu, ``Optimal power allocation for secure directional modulation
  networks with a full-duplex uav user,'' \emph{Sci. China Inf. Sci.}, vol.~62,
  no. 008, pp. 1--12, 2019.

\bibitem{book2009}
P.~Antonik, ``An investigation of a frequency diverse array,'' \emph{Dept.
  Electron. Elect. Eng., Univ. College London, London, U.K.}, 2009.

\bibitem{Shu2020DM}
F.~{Shu}, T.~{Shen}, L.~{Xu}, Y.~{Qin}, S.~{Wan}, S.~{Jin}, X.~{You}, and
  J.~{Wang}, ``Directional modulation: A physical-layer security solution to
  {B5G} and future wireless networks,'' \emph{IEEE Network}, vol.~34, no.~2,
  pp. 210--216, 2020.

\bibitem{Qin2018}
F.~{Shu}, Y.~{Qin}, T.~{Liu}, L.~{Gui}, Y.~{Zhang}, J.~{Li}, and Z.~{Han},
  ``Low-complexity and high-resolution {DOA} estimation for hybrid analog and
  digital massive mimo receive array,'' \emph{IEEE Trans. Commun.}, vol.~66,
  no.~6, pp. 2487--2501, June 2018.

\bibitem{Han1}
H.~{Wu}, X.~{Tao}, Z.~{Han}, N.~{Li}, and J.~{Xu}, ``Secure transmission in
  misome wiretap channel with multiple assisting jammers: Maximum secrecy rate
  and optimal power allocation,'' \emph{IEEE Trans. Commun.}, vol.~65, no.~2,
  pp. 775--789, Feb. 2017.

\bibitem{Sammartino2013}
P.~F. Sammartino, C.J.Baker, and H.D.Griffiths, ``Frequency diverse mimo
  techniques for radar,'' \emph{IEEE Trans.Aerosp.Electron.Syst.}, vol.~49,
  no.~1, pp. 201--222, Jan. 2013.

\bibitem{Wang2015}
W.~Q. Wang, ``Frequency diverse array antenna: New opportunities,'' \emph{IEEE
  Antennas Propag.Mag}, vol.~57, no.~2, pp. 145--152, Apr. 2015.

\bibitem{liu2017random}
Y.~Liu, H.~Ruan, L.~Wang, and A.~Nehorai, ``The random frequency diverse array:
  A new antenna structure for uncoupled direction-range indication in active
  sensing,'' \emph{IEEE J. Sel. Top. Sign. Proces.}, vol.~11, no.~2, pp.
  295--308, 2017.

\bibitem{Hu2017SPWT}
J.~Hu, S.~Yan, F.~Shu, J.~Wang, J.~Li, and Y.~Zhang, ``Artificial-noise-aided
  secure transmission with directional modulation based on random frequency
  diverse arrays,'' \emph{IEEE Access}, vol.~5, pp. 1658 -- 1667, Jan. 2017.

\bibitem{Shu2018SPWT}
F.~Shu, X.~Wu, J.~Hu, J.~Li, R.~Chen, and J.~Wang, ``Secure and precise
  wireless transmission for random-subcarrier-selection-based directional
  modulation transmit antenna array,'' \emph{IEEE J. Sel. Areas Commun},
  vol.~36, no.~4, pp. 890 -- 904, Apr. 2018.

\bibitem{Shen2019}
T.~{Shen}, S.~{Zhang}, R.~{Chen}, J.~{Wang}, J.~{Hu}, F.~{Shu}, and J.~{Wang},
  ``Two practical random-subcarrier-selection methods for secure precise
  wireless transmissions,'' \emph{IEEE Trans. Veh. Technol.}, vol.~68, no.~9,
  pp. 9018--9028, Sep. 2019.

\bibitem{Zhu2009Chunk}
H.~Zhu and J.~Wang, ``Chunk-based resource allocation in {OFDMA} systems - part
  i: Chunk allocation,'' \emph{IEEE Trans. Commun.}, vol.~57, no.~9, pp.
  2734--2744, Sep. 2009.

\bibitem{Zhu2012Chunk}
------, ``Chunk-based resource allocation in {OFDMA} systems¡ªpart ii: Joint
  chunk, power and bit allocation,'' \emph{IEEE Trans. Commun.}, vol.~60,
  no.~2, pp. 499--509, Feb. 2012.

\bibitem{Shen2020}
T.~{Shen}, Y.~{Lin}, J.~{Zou}, Y.~{Wu}, F.~{Shu}, and J.~{Wang},
  ``Low-complexity leakage-based secure precise wireless transmission with
  hybrid beamforming,'' \emph{IEEE Wireless Commun. Lett.}, vol.~9, no.~10, pp.
  1687 -- 1691, Oct. 2020.

\bibitem{2017IRS}
F.~R. S.~Hu and O.~Edfors, ``Beyond massive mimo: The potential of data
  transmission with large intelligent surfaces,'' \emph{IEEE Trans. Signal
  Process.}, vol.~66, no.~10, pp. 2746--2758, 2017.

\bibitem{panCM2021}
C.~Pan, H.~Ren, K.~Wang, J.~F. Kolb, M.~Elkashlan, M.~Chen, M.~Di~Renzo,
  Y.~Hao, J.~Wang, A.~L. Swindlehurst, X.~You, and L.~Hanzo, ``Reconfigurable
  intelligent surfaces for 6g systems: Principles, applications, and research
  directions,'' \emph{IEEE Communications Magazine}, vol.~59, no.~6, pp.
  14--20, 2021.

\bibitem{Wu2019}
Q.~{Wu} and R.~{Zhang}, ``Intelligent reflecting surface enhanced wireless
  network via joint active and passive beamforming,'' \emph{IEEE Trans. Wirel.
  Commun.}, vol.~18, no.~11, pp. 5394--5409, Nov. 2019.

\bibitem{Wu2020Tcom}
------, ``Beamforming optimization for wireless network aided by intelligent
  reflecting surface with discrete phase shifts,'' \emph{IEEE Trans. Commun.},
  vol.~68, no.~3, pp. 1838--1851, Mar. 2020.

\bibitem{Wu2020WCL}
------, ``Weighted sum power maximization for intelligent reflecting surface
  aided swipt,'' \emph{IEEE Wireless Commun. Lett.}, vol.~9, no.~5, pp.
  586--590, May 2020.

\bibitem{Jiang2020}
W.~{Jiang}, Y.~{Zhang}, J.~{Wu}, W.~{Feng}, and Y.~{Jin}, ``Intelligent
  reflecting surface assisted secure wireless communications with
  multiple-transmit and multiple-receive antennas,'' \emph{IEEE Access},
  vol.~8, pp. 86\,659--86\,673, May 2020.

\bibitem{CVX2004}
S.~Boyd and L.~Vandenberghe.\hskip 1em plus 0.5em minus 0.4em\relax U.K.:
  Cambridge Univ. Press, 2004.

\bibitem{book2012}
A.~Ben-Tal and A.~Nemirovski, \emph{Lectures on Modern Convex Optimization:
  Analysis, Algorithms, and Engineering Applications}, 01 2012.

\bibitem{Horn1985Matrix}
R.~A. Horn and C.~R. Johnson, \emph{Matrix Analysis}, Cambridge, U.K.:
  Cambridge Univ. Press, 1987.

\end{thebibliography}

\end{document}